\documentclass[12pt,thmsa]{article}
\usepackage{amsmath}
\usepackage{amssymb}

\input tcilatex

\begin{document}

\begin{center}
\bigskip \emph{\ Self-Creation Cosmology - A Review}

Garth A. Barber

The Vicarage, Woodland Way, Tadworth, Surrey, England KT206NW

Tel: +44 01737 832385 \qquad

e-mail: garth.barber@virgin.net

\allowbreak \textit{Abstract}
\end{center}

Over 60 authors have worked on various versions of self-creation cosmology
(SCC) since the original paper in 1982. These papers adapted the Brans Dicke theory to create mass out of the universe's self contained scalar, gravitational and matter fields. The most recent 2002 version of the theory was concordant with all previous standard tests of GR but was falsified by the Gravity Probe B geodetic precession experiment. Here the different versions of the theory are reviewed and it is noted that the 2002 theory not only reduces to the second 1982 theory when cast into a the 'true' form of the scalar field stress-energy tensor but also then passes the GP-B test. Further experiments are able to resolve the SCC-GR degeneracy, one of which is briefly described. SCC may be able to explain some intriguing anomalies also, in the spherically symmetric
One-Body problem, it transpires that the temporal Newtonian potential is three times larger than the spatial one. Cosmological solutions to the field equations, which have been published elsewhere, are here extended to show that the cosmological density parameter Omega =1.

\section{A r\'{e}sum\'{e} of SCC papers}

All self-creation cosmology (SCC) papers (Barber, 1982, 2002) modify the
Brans Dicke theory (Brans \& Dicke, 1961) to include mass creation. The
first paper (Barber, 2002) suggested two toy models, the first of which was
rejected in that paper on the grounds of a gross violation of the
equivalence principle and experiment while Brans (Brans, 1987) subsequently
showed that it was also internally inconsistent. The second theory (SCC2),
however, has led to over 80 citations. A more comprehensive version was
published (Barber, 2002), that was unfortunately cast in a form in which its
prediction of geodesic precession (Barber, 2006) was subsequently falsified
by the Gravity Probe B (GPB) experiment. This falsification has stimulated
further work. Here a corrected form is presented which fortuitously proves
to be a version of the second 1982 theory. It not only passes all
experimental tests to date (now including GPB) but also makes an interesting
cosmological model.

Many authors have tried to include a scalar field in General Relativity
(GR), not only to more fully include Mach's Principle (such as in BD), but
also to accommodate certain attempts at quantum gravity in which scalar
tensor theories that have proven to describe the low energy limit.
Unfortunately, one persistent problem has been the perturbation of the
metric by the addition of a\ scalar field consequentially the experimental
agreement with GR has been lost. As a consequence only a very weak field in
such theories has been thought to be empirically viable, at least within the
Solar System laboratory. This problem is overcome in SCC where the effect of
a non-minimally connected scalar field exactly compensates for the
perturbation of space-time, thus restoring its predictions to those of GR.

This non-minimal connection of the\ scalar field to matter in\ the SCC
Jordan frame (JF) identifies it in Quiros' classification scheme as a class
II scalar-tensor theory (Quiros, 1999). A special\ feature of the theory is
the JF locally conserves energy (measured in the 'Machian', i.e. Centre of
Mass, frame of reference) whereas the Einstein frame (EF) conserves
energy-momentum. This requirement, together with the principle of mutual
interaction (see below), determines the coupling constant to be $\omega =-%
\frac{3}{2}$. As a consequence the EF of the theory proves to be canonical GR%
\textit{\ in vacuo} . 

The question as to which frame is physical, that is that which relates to
experimental measurement, depends in this theory on the physical standard
chosen to measure time. On the one hand, in the EF, where atomic masses are
constant, the time standard is set by an atomic clock and a second may be
defined conventionally as the duration of exactly 9.19263177x10$^{9}$
periods of the radiation emitted by the transition between the two hyperfine
levels of the ground state of the caesium 133 atom. On the other hand, in
the JF, a second may be defined as the duration of exactly 1.604.10$^{11}$
periods of the radiation corresponding to the peak of the isotropic CMB
black body spectrum. As both systems of time measurement are physically
significant both definitions are 'physical' in an experimental sense. Time
is the fundamental measurement in both frames, determined by Bohr atomic
frequencies in the EF and the frequency of a reference photon, carefully
defined, in the JF. By definition the speed of light \textit{in vacuo }is
invariant in both. The two frames are synchronized at some local event in
the present epoch used to set laboratory standards but they will generally
diverge at other times and locations. It is the contention of this theory
that gravitational, and hence cosmological, problems have to be solved in
the JF and this is used throughout unless specifically stated otherwise.
Consequently ephemeris time is to be defined in the JF and suffers a secular
clock drift relative to time measured by atomic clocks.

The JF version of the theory has been described as 'semi-metric' because
although there is a metric and photons do follow the geodesics of that
frame, freely falling test particles do not. \textit{In vacuo} an extra
force acts on 'free falling particles', compensates for the perturbation of
the GR metric. This results in SCC test particles following canonical GR
geodesics in both frames and SCC and GR predictions are identical in all
tests to date. Nevertheless, there are two further experiments that will be
able to distinguish between them. In addition the theory also offers
explanations for the Pioneer Anomaly and some other intriguing anomalies
(Barber, 2002b).

\section{Deriving the field equations}

Following BD, SCC theories incorporate Mach's Principle (MP) by assuming the
inertial masses of fundamental particles are dependent upon their
interaction with a scalar field $\phi \approx \frac{1}{G_{N}}$ coupled to
the large scale distribution of matter in motion. This coupling is described
by a field equation of the simplest general covariant form: 
\begin{equation}
\Box \phi =4\pi \lambda T_{M}^{\;}\text{ ,}  \label{eq.1}
\end{equation}%
where $T_{M\;}^{\;}$ is the trace, ($T_{M\;\sigma }^{\;\;\sigma }$), of the
energy momentum tensor describing all non-gravitational and non-scalar field
energy and $\lambda $ is some undetermined coupling constant of the order
unity. In the spherically symmetric One Body problem of BD 
\begin{equation}
\underset{r\rightarrow \infty }{Lim}\phi \left( r\right) =\frac{\psi }{G_{N}}%
\text{ ,}  \label{eq.2}
\end{equation}%
where $\psi $ is of the order unity and determined by $\lambda $. (see
equation \ref{eq.135} below). The gravitational field equation included the
energy-momentum tensor of the scalar field energy $T_{\phi \,\mu \nu }$
where $T_{M\mu \nu }$ and $T_{\phi \,\mu \nu }$ are the energy momentum
tensors describing the matter and scalar fields respectively.

In both GR and BD the equation describing the interchange of energy between
matter and gravitation is,

\begin{equation}
\nabla _{\mu }T_{M\;\nu }^{\;\mu }=0\text{ ,}  \label{eq.3}
\end{equation}

This equation, which conserves four-momentum, is a consequence of the
equivalence principle, and in the theory of BD it guarantees that the scalar
field interacts with matter only by adapting the curvature of space-time and
in no other way, i.e. ordinary matter is minimally coupled to the scalar
field. In BD and SCC the scalar field is included in the gravitational field
equation which becomes%
\begin{equation}
R_{\mu \nu }-\frac{1}{2}g_{\mu \nu }R=\frac{8\pi }{\phi }\left[ T_{M\mu \nu
}+T_{\phi \,\mu \nu }\right] \text{ .}  \label{eq.4}
\end{equation}

In the 1982 SCC2 theory the scalar field was minimally coupled to the metric
and therefore only interacted with the material universe by determining the
gravitational coefficient $G$ with a field equation 
\begin{equation}
R_{\mu \nu }-\frac{1}{2}g_{\mu \nu }R=\frac{8\pi }{\phi }T_{M\mu \nu }\text{
.}  \label{eq.5}
\end{equation}%
As $\phi $ was not constant then $\nabla _{\mu }T_{M\;\nu }^{\;\mu }\neq 0$
and ordinary matter was non-minimally connected to the scalar field. However
this hypothesis was criticized by Brans, (Brans, 1987), on the basis of the
difficulty of defining a metric if the paths of photons are not
null-geodesics. Nevertheless SCC2 continues to provoke discussion. [A
selection from 81 'other-author' papers is given below].

\subsubsection{Two principles}

\paragraph{The principle of mutual interaction}

What is to constrain the process if mass is indeed created out of
gravitational and scalar fields? The SCC answer is the postulate of the
Principle of Mutual Interaction (PMI) (Barber, 2002), which states that:
"The scalar field is a source for the matter-energy field if and only if the
matter-energy field is a source for the scalar field." In more specific
terms, if the source for the scalar field is the trace of the matter
stress-energy tensor then the divergence of the matter stress-energy tensor
should be coupled to this trace. 

\bigskip\ The conservation equation \ref{eq.3} is consequently replaced in
SCC with the PMI 'creation' equation of the form 
\begin{equation}
\nabla _{\mu }T_{M\;\nu }^{.\;\mu }=f_{\nu }\left( \phi \right) \Box \phi
=4\pi \lambda f_{\nu }\left( \phi \right) T_{M\;}^{\;}  \label{eq.6}
\end{equation}%
As a consequence relativistic energy such as light, which is trace free,
still obeys the Equivalence Principle and Brans' latter criticism is
resolved because, at least \textit{in vacuo,} \textit{\ } 
\begin{equation}
\nabla _{\mu }T_{em\quad \nu }^{\quad \mu }=4\pi \lambda f_{\nu }\left( \phi
\right) T_{em}^{\;}=4\pi \lambda f_{\nu }\left( \phi \right) \left(
3p_{em}-\rho _{em}\right) =0  \label{eq.7}
\end{equation}%
where $p_{em}$ and $\rho _{em}$ are the pressure and density of an
electromagnetic radiation field with an energy momentum tensor $T_{em\,\mu
\nu }$ and where $p_{em}=\frac{1}{3}\rho _{em}$. \ So although the
equivalence principle is violated in general it is not so for photons, which
still travel through empty space on (null) geodesic paths. Therefore,
although the theory is not fully metric in the classical sense, as photons
still do obey the equivalence principle it might be called a semi-metric
theory. On the other hand particles with mass suffer a 'scalar field force'
perturbing their trajectories from geodesics. In other words, ordinary
matter and relativistic energy are non-minimally and minimally coupled to
the scalar field respectively.

\paragraph{The local conservation of energy}

A second principle, that of the principle of the local conservation of
energy complements equation \ref{eq.6} and fully determines the theory. This
postulates that the potential energy expended in moving an object in a
gravitational field should translate into an increase in rest mass. If $\Phi
_{N}\left( x^{\mu }\right) $ is the dimensionless Newtonian gravitational
potential defined by a measurement of acceleration in a local experiment in
a frame of reference co-moving with the Centre of Mass frame (CoM), with

\begin{equation}
\frac{d^{2}r}{dt^{2}}=-\nabla \Phi _{N}\left( r\right)  \label{eq.8}
\end{equation}

and normalized so that $\Phi _{N}\left( \infty \right) $ $=0$ , then%
\begin{equation}
\frac{1}{m_{p}\left( x^{\mu }\right) }\mathbf{\nabla }m_{p}\left( x^{\mu
}\right) =\mathbf{\nabla }\Phi _{N}\left( x^{\mu }\right) ,  \label{eq.9}
\end{equation}%
where $m_{p}(x^{\mu })$ is measured locally at $x^{\mu }$ and $c=1$
throughout. This has the solution%
\begin{equation}
m_{p}(x^{\mu })=m_{0}\exp [\Phi _{N}\left( x^{\mu }\right) ],  \label{eq.10}
\end{equation}%
where%
\begin{equation*}
\qquad m_{p}\left( r\right) \rightarrow m_{0}\qquad as\qquad r\rightarrow
\infty .
\end{equation*}

This can be seen by considering time dilation observed as the gravitational
red shift of light.

\paragraph{The gravitational red-shift of light}

The gravitational red shift of light is now to considered in order to
examine the measurement problem in both the EF and the JF. This analysis
depends on the assumption that if no work is done on, or by, a projectile
while in free fall then its energy $E$ , $P^{0}$ , is conserved \textit{when
measured in a specific frame of reference}, that of the CoM of the system.%
\textit{\ }In a gedanken, 'thought', experiment, construct a laboratory at
the co-moving centroid, the CoM, of the system. Connect it to the outside
world by a radial tube through which identical test masses and photons may
be projected \textit{in vacuo}. Launch such projectiles, with rest masses, $%
m_{0}$, at the CoM at various velocities to reach maximum altitudes $r_{i}$
where $r_{i}$ varies increasing from $R$, the radius of the central mass, to
infinity. The 'rest' mass of the projectile $m_{c}\left( r\right) $ , the
'coordinate' mass, is in general to be a function of altitude measured in
the CoM frame of reference.

First consider such a photon emitted by one atom at altitude $x_{2}$ and
absorbed by another at an altitude $x_{1}$. The emission and absorption
frequencies of the photon, $\nu \left( x_{2}\right) $ and $\nu \left(
x_{1}\right) $, are determined by comparing the arrival times of two
adjacent wave fronts emitted from one point in a gravitational field at $%
\left( x_{2}\right) $ and received at another at $\left( x_{1}\right) $. The
standard time dilation relationship is thereby derived 
\begin{equation}
\frac{\nu \left( x_{2}\right) }{\nu \left( x_{1}\right) }=\left[ \frac{%
-g_{00}\left( x_{2}\right) }{-g_{00}\left( x_{1}\right) }\right] ^{\frac{1}{2%
}}\text{ .}  \label{eq.46}
\end{equation}%
Hence substituting $x_{2}=r$ and $x_{1}=\infty $ in equation \ref{eq.46},
where $g_{00}\left( x_{1}\right) =-1$, and writing $\nu \left( \infty
\right) $ as $\nu _{0}$, yields the standard (GR) gravitational red shift
relationship 
\begin{equation}
\nu \left( r\right) =\nu _{0}\left[ -g_{00}\left( r\right) \right] ^{\frac{1%
}{2}}\text{ ,}  \label{eq.47}
\end{equation}%
where the observer is at infinite altitude receiving a photon emitted at
altitude $r$.

Now consider the various projectiles. With the standard definition of proper
time $\tau $ from the metric 
\begin{equation}
d\tau ^{2}=-g_{\mu \nu }dx^{\mu }dx^{\nu }\text{ .}  \label{eq.48}
\end{equation}%
The 4-momentum vector of the projectile is defined 
\begin{equation}
P^{\mu }=m_{c}\frac{dx^{\mu }}{d\tau }\text{ .}  \label{eq.49}
\end{equation}%
The time component of 4-momentum $P^{\mu }$ is the total 'relative' energy $%
E $ and the space components form the 'relative' 3-momentum $\underline{p}$ .

Now from equation \ref{eq.48} we obtain 
\begin{equation}
\frac{d\tau ^{2}}{dt^{2}}=-g_{00}-2g_{i0}v^{i}-v^{2}\text{ ,}  \label{eq.50}
\end{equation}%
\begin{equation}
\text{where }v^{i}=\frac{dx^{i}}{dt}\text{ and }v^{2}=g_{ij}\frac{dx^{i}}{dt}%
\frac{dx^{j}}{dt}\text{ .}  \label{eq.51}
\end{equation}%
Therefore in a spherically symmetric, non-rotating, metric with $g_{i0}=0$, 
\begin{equation}
-g_{00}E^{2}=m_{c}^{2}+\underline{p}^{2}\text{ .}  \label{eq.52}
\end{equation}%
This is the spherically symmetric curved space-time equivalent to the SR
identity 
\begin{equation}
E^{2}=m_{c}^{2}+\underline{p}^{2}\text{ .}  \label{eq.53}
\end{equation}

Now consider two of the projectiles as they momentarily reach their
respective apocentres at maximum altitude $r$, and $r+\delta r$. As they are
momentarily stationary in the CoM frame $\underline{p}=0$ . The difference
between the two adjacent projectiles at their apocentres is that one has a
total energy and rest mass of $E\left( r\right) $, and $m_{c}\left( r\right) 
$, and the other $E\left( r+\delta r\right) $, and $m_{c}\left( r+\delta
r\right) $ . Expanding for small $\delta r$, and where a prime ($^{\prime }$%
) means $\frac{d}{dr}$, in the limit $\delta r\rightarrow 0$ we obtain 
\begin{equation}
\frac{1}{2}\frac{\left[ -g_{00}^{\prime }\left( r\right) \right] }{\left[
-g_{00}\left( r\right) \right] }+\frac{E^{\prime }\left( r\right) }{E\left(
r\right) }=\frac{m_{c}^{\prime }\left( r\right) }{m_{c}\left( r\right) }%
\text{ .}  \label{eq.54}
\end{equation}%
Here two identical projectiles are compared which are separated by an
infinitesimal increase in altitude. The only difference between them is the
infinitesimal energy $\delta E$ required to raise such a projectile from $r$
to $r+\delta r$. Although the Newtonian potential $\Phi _{N}\left( r\right) $
is defined by 
\begin{equation}
\nabla ^{2}\Phi _{N}\left( r\right) =4\pi G_{N}\overset{0}{T}^{00}=4\pi
G_{N}\rho  \label{eq.55}
\end{equation}%
which is normalized, 
\begin{equation*}
\Phi _{N}\left( \infty \right) =0\text{ ,}
\end{equation*}%
it is actually measured in a Cavendish type laboratory experiment by the
force vector acting on a body, which is given by 
\begin{equation}
\mathbf{F}=-m_{p}\mathbf{\nabla }\Phi _{N}\left( r\right) \text{ .}
\label{eq.56}
\end{equation}%
Then if the mass $m\left( r\right) $ is raised a height $\delta \mathbf{r}$
against this force, the infinitesimal energy $\delta E$ required is 
\begin{equation}
\delta E=-\mathbf{F}\circ \delta \mathbf{r}=m_{p}\left( r\right) \mathbf{%
\nabla }\Phi _{N}\left( r\right) \circ \delta \mathbf{r}\text{ .}
\label{eq.57}
\end{equation}%
That is in the radial case 
\begin{equation}
\delta E=m_{p}\left( r\right) \Phi _{N}^{\prime }\delta r\text{ ,}
\label{eq.58}
\end{equation}%
where $m_{p}\left( r\right) $ is that physical mass entering into the
Newtonian gravitational equation. Define such physical mass, momentarily at
rest, as 
\begin{equation}
m_{p}\left( r\right) =E\left( r\right) \text{ ,}  \label{eq.59}
\end{equation}%
so that the total \textquotedblright relative\textquotedblright\ energy at
an altitude $r$ is its rest mass at that altitude, measured in the CoM frame
of reference. In the limit $\delta r\rightarrow 0$ equations \ref{eq.58} and %
\ref{eq.59} become 
\begin{equation}
\frac{E^{\prime }(r)}{E\left( r\right) }=\Phi _{N}^{\prime }\left( r\right) 
\text{ ,}  \label{eq.60}
\end{equation}%
which when substituted in equation \ref{eq.54} yields 
\begin{equation}
\frac{1}{2}\frac{\left[ -g_{00}^{\prime }\left( r\right) \right] }{\left[
-g_{00}\left( r\right) \right] }+\Phi _{N}^{\prime }\left( r\right) =\frac{%
m_{c}^{\prime }\left( r\right) }{m_{c}\left( r\right) }\text{ . }
\label{eq.61}
\end{equation}%
This integrates directly, 
\begin{equation}
\frac{1}{2}\ln \left[ -g_{00}\left( r\right) \right] +\Phi _{N}\left(
r\right) =\ln \left[ m\left( r\right) \right] +k  \label{eq.62}
\end{equation}%
where $k$ is determined in the limit $r\rightarrow \infty $, $g_{00}\left(
r\right) \rightarrow -1$, $\Phi _{N}\left( r\right) \rightarrow 0$ and $%
m\left( r\right) \rightarrow m_{0}$. The rest mass, $m\left( r\right) $, of
a projectile at altitude $r$, evaluated in the co-moving CoM frame is
therefore given by 
\begin{equation}
m_{c}\left( r\right) =m_{0}\exp \left[ \Phi _{N}\left( r\right) \right] %
\left[ -g_{00}\left( r\right) \right] ^{\frac{1}{2}}\text{ .}  \label{eq.63}
\end{equation}%
This is the value, $m_{c}\left( r\right) $, given by an observer at infinite
altitude, where Special Relativity and a ground state solution to the theory
are recovered, with well defined particle rest mass $m_{0}$, 'looking down'
to a similar particle at an altitude $r$. From this expression it is obvious
that with our assumption of the conservation of energy, $P^{0}$ , in the CoM
frame gravitational time dilation, the factor $\left[ -g_{00}\left( r\right) %
\right] ^{\frac{1}{2}}$, applies to massive particles as well as to photons.
As physical experiments measuring the frequency of a photon compare its
energy with the mass of the atom it interacts with, it is necessary to
compare the masses (defined by equation \ref{eq.63}) of two atoms at
altitude, $r$ and $\infty $, with the energy (given by equation \ref{eq.47})
of a \textquotedblright reference\textquotedblright\ photon transmitted
between them. This yields the physical rest mass $m_{p}\left( r\right) $ as
a function of altitude 
\begin{equation}
\frac{m_{p}\left( r\right) }{\nu \left( r\right) }=\frac{m_{0}}{\nu _{0}}%
\exp \left[ \Phi _{N}\left( r\right) \right] \text{ .}  \label{eq.64}
\end{equation}

Equation \ref{eq.64} is a result relating observable quantities, but how is
it to be interpreted? In other words how are mass and frequency to be
measured in any particular frame? In the GR EF (and BD JF) the physical rest
mass of the atom is defined to be constant, hence prescribing ($\widetilde{x}%
^{\mu }$), with $m_{p}\left( \widetilde{r}\right) =m_{0}$. In this case
equation \ref{eq.64} becomes 
\begin{equation}
\nu \left( \widetilde{r}\right) =\nu _{0}\left( 1-\widetilde{\Phi }%
_{N}\left( \widetilde{r}\right) +...\right) \text{ .}  \label{eq.65}
\end{equation}%
Hence photons transmitted out of a gravitational potential well are said to
exhibit a red shift which is equal to the dimensionless Newtonian potential $%
\widetilde{\Phi }_{N}$, and equal in GR, \textquotedblright
coincidentally\textquotedblright , to the time dilation effect, the factor$%
\left[ -\widetilde{g}_{00}\left( \widetilde{r}\right) \right] ^{\frac{1}{2}}$%
. That is, compared to reference atoms they mysteriously appear to lose
(potential) energy.

However in the SCC JF rest mass is given by the expression equation \ref%
{eq.10}, consequently a comparison of equation \ref{eq.64} with the equation
for rest mass in this frame yields 
\begin{equation}
\nu \left( r\right) =\nu _{0}\text{ .}  \label{eq.66}
\end{equation}

Therefore in the SCC JF, in which energy is locally conserved, gravitational
red shift is interpreted not as a loss of potential energy by the photon but
rather as a gain of potential energy by the apparatus measuring it. Time
dilation is subsumed into the change of mass of the observer's apparatus.
The 'absolute' time of the JF is defined in a specific or 'preferred' frame
of reference, which is that one 'anchored' to the co-moving centroid, or
centre-of-mass of the system.

It is important to note that in this frame the frequency, and hence
wavelength and energy, of a free photon is invariant, even when traversing
curved space-time.

On the other hand, as experiments using physical apparatus refer
measurements of energy and mass to the mass of the atoms of which they are
composed, such observations interpret rest masses to be constant by
definition. In SCC such experiments are analyised in the EF in which
equation \ref{eq.65} describes gravitational red shift. Using either frame
the gravitational red shift prediction in SCC is in agreement with GR and
all observations to date.

Because physical rulers and clocks vary with atomic masses in the JF SCC
interprets physical observations by using light to fulfill the fundamental
role of measuring the universe. This may be seen to be a similar, but more
general, method used in E.A.Milne's theory of Kinematic Relativity. (Milne,\
1935, 1948).

\subsection{The SCC conformal frames of reference}

Weyl's hypothesis, (Weyl, 1918), led to the concept that the space-time
manifold $M$ is not equipped with a unique metric as in GR but a class [$%
g_{\mu \nu }$] of conformally equivalent Lorentz metrics $g_{\mu \nu }$. In
a conformal transformation one metric transforms into a physically
equivalent alternative according to 
\begin{equation}
g_{\mu \nu }\rightarrow \widetilde{g}_{\mu \nu }=\Omega ^{2}g_{\mu \nu }%
\text{ .}  \label{eq.11}
\end{equation}

The self creation, ( $\nabla _{\mu }T_{M\;\nu }^{\;\mu }\neq 0$ ), of SCC
requires the JF scalar field to be non-minimally connected to matter, hence
it is an example of the work of Magnano and Sokolowski (Magnano \&
Sokolowski, 1994) who applied conformal duality to GR in order to include a
scalar field as an additional source of gravity. In their case [in contrast
to BD, (Dicke, 1962)] ordinary matter is non-minimally coupled to the scalar
field in the JF and it is minimally coupled in the EF. In the JF particle
masses and the Gravitational 'constant' vary, whereas in the EF they are
both constant.

The Lagrangian density in the JF is given by

\begin{equation}
L^{SCC}[g,\phi ]=\frac{\sqrt{-g}}{16\pi }\left( \phi R-\frac{\omega }{\phi }%
g^{\mu \nu }\nabla _{\mu }\phi \nabla _{\nu }\phi \right) +L_{matter}[g,\phi
]\text{ ,}  \label{eq.12}
\end{equation}

\bigskip which, on varying the metric components produces the gravitational
field equation,

\begin{eqnarray}
R_{\mu \nu }-\frac{1}{2}g_{\mu \nu }R &=&\frac{8\pi }{\phi }T_{M\mu \nu }+%
\frac{\omega }{\phi ^{2}}\left( \nabla _{\mu }\phi \nabla _{\nu }\phi -\frac{%
1}{2}g_{\mu \nu }\nabla _{\sigma }\phi g^{\mu \sigma }\nabla _{\sigma }\phi
\right)  \label{eq.12a} \\
&&+\frac{1}{\phi }\left( \nabla _{\mu }\nabla _{\nu }\phi -g_{\mu \nu }\Box
\phi \right) \text{ ,}  \notag
\end{eqnarray}

and the conformal dual of equation \ref{eq.12} is, (Dicke, 1962)%
\begin{eqnarray}
L^{SCC}[\widetilde{g},\widetilde{\phi }] &=&\frac{\sqrt{-\widetilde{g}}}{%
16\pi }\left[ \widetilde{\phi }\widetilde{R}+6\widetilde{\phi }\widetilde{%
\Box }\ln \Omega \right] +\widetilde{L}_{matter}^{SCC}[\widetilde{g},%
\widetilde{\phi }]  \label{eq.13} \\
&&-\frac{\sqrt{-\widetilde{g}}}{8\pi }\left( 2\omega +3\right) \frac{%
\widetilde{g}^{\mu \nu }\widetilde{\nabla }_{\mu }\Omega \widetilde{\nabla }%
_{\nu }\Omega }{\Omega ^{2}} \\
&&\ -\frac{\sqrt{-\widetilde{g}}}{16\pi }\omega \left[ 4\frac{\widetilde{g}%
^{\mu \nu }\widetilde{\nabla }_{\mu }\Omega \widetilde{\nabla }_{\nu }%
\widetilde{\phi }}{\Omega }+\frac{\widetilde{g}^{\mu \nu }\widetilde{\nabla }%
_{\mu }\widetilde{\phi }\widetilde{\nabla }_{\nu }\widetilde{\phi }}{%
\widetilde{\phi }}\right] \text{ .}  \notag
\end{eqnarray}

As mass is conformally transformed according to

\begin{equation}
m\left( x^{\mu }\right) =\Omega \widetilde{m}_{0}  \label{eq.14}
\end{equation}

(see Dicke, 1962), where $m\left( x^{\mu }\right) $ is the mass of a
fundamental particle in the JF and $\widetilde{m}_{0}$ its invariant mass in
the EF then equations \ref{eq.10} and \ref{eq.14} require 
\begin{equation}
\Omega =\exp \left[ \Phi _{N}\left( x^{\mu }\right) \right] \text{ .}
\label{eq.15}
\end{equation}

The metrics thus relate \textit{in vacuo} according to equation \ref{eq.11} 
\begin{equation}
g_{\mu \nu }\rightarrow \widetilde{g}_{\mu \nu }=\exp \left[ 2\Phi
_{N}\left( x^{\mu }\right) \right] g_{\mu \nu }\text{ ,}  \label{eq.16}
\end{equation}%
where $\widetilde{g}_{\mu \nu }$ is the GR metric. Compare this to Nordstr%
\"{o}m's pre-GR attempt to develop a relativistic gravitational theory by a
conformal mapping of the Minkowski metric $\eta _{\mu \nu }$, 
\begin{equation}
\eta _{\mu \nu }\rightarrow g_{\mu \nu }=\exp \left[ 2\Phi _{N}\left( x^{\mu
}\right) \right] \eta _{\mu \nu }\text{ .}  \label{eq.17}
\end{equation}%
(Nordstr\"{o}m, 1913, see also review by Brans, 1997). Whereas Nordstr\"{o}m
tried to include gravitational potential energy in Special Relativity (SR)
by conformally mapping $\eta _{\mu \nu }$ the SR metric, in SCC it is
included by a conformal mapping onto $\widetilde{g}_{\mu \nu }$ the GR
metric. However this mapping is only exact \textit{in vacuo}, therefore SCC
is not a simple conformal mapping of GR. There is a significant physical
difference which will reveal itself as the theory develops, although the
vacuum solutions of geodesic orbits are the same in both theories. Note also
the dimensionless $Gm^{2}$ will not be invariant under translation within a
gravitational field because of the real variation of mass caused by such
inclusion of potential energy.

\paragraph{The transformation of $\protect\phi $}

The question how $\phi $ transforms\ has to be addressed if SCC is to
concatenate potential energy within the definition of inertial mass. In
scalar-tensor theories the conformal transformation of the scalar field is
assumed to depend on the dimensionless and therefore invariant,%
\begin{equation}
Gm^{2}=\widetilde{G}\widetilde{m}^{2}  \label{eq.18}
\end{equation}%
so by equations \ref{eq.2} and \ref{eq.14} we have 
\begin{equation}
\text{ }\widetilde{\phi }=\phi \Omega ^{-2}\text{ .}  \label{eq.19}
\end{equation}%
So, if the conformal transformation $\Omega $ is defined by 
\begin{equation}
\Omega =\left( G\phi \right) ^{\alpha }  \label{eq.20}
\end{equation}%
then 
\begin{equation}
\widetilde{\phi }=G^{-2\alpha }\phi ^{\left( 1-2\alpha \right) }\text{ ,}
\label{eq.21}
\end{equation}%
which, as $\widetilde{\phi }$ is a constant in BD, determines $\alpha =\frac{%
1}{2}$ and therefore

\begin{equation}
\Omega =\sqrt{\left( G\phi \right) }\text{.}  \label{eq.22}
\end{equation}

This relationship in standard scalar-tensor theory depends on the assumption
that it is $Gm^{2}$ that is invariant under the transformation and that then
completely determines the conformal factor $\Omega $. However in the SCC JF
the local conservation of energy requires a different form for $\Omega $, is
this in fact possible? In responding to this question we note that in this
frame potential energy is to be convoluted with inertial mass and hence
gravitation, therefore it is reasonable to assume the dimensionless
invariant has to be enlarged to include such potential energy in the form of
the dimensionless Newtonian potential $\Phi _{N}\left( x^{\mu }\right) $.
The general conformal invariant therefore becomes 
\begin{equation}
Gm^{2}f\left( \Phi _{N}\right) =\widetilde{G}\widetilde{m}^{2}  \label{eq.23}
\end{equation}%
The form of $f\left( \Phi _{N}\right) $ is determined by the exact
relationship between $G\phi \left( x_{\mu }\right) $ and $\Phi _{N}\left(
x_{\mu }\right) $.

[For example, in the static dust filled universe where $\nabla ^{2}\phi
=-4\pi \lambda \rho $, and $\nabla ^{2}\Phi _{N}=-4\pi G\rho $ the
dimensionless $f\left( \Phi _{N}\right) $ is determined to be (when $\lambda
=1$ as below)%
\begin{equation*}
f(\Phi _{N})=G_{N}\phi \exp \left( 2G_{N}\phi \right) \text{.]}
\end{equation*}%
In the EF frame $\widetilde{m}\left( x^{\mu }\right) $ is constant,
consequently so are $\widetilde{G}$ and hence $\widetilde{\phi }$ and in
this case 
\begin{equation}
\nabla _{\alpha }\widetilde{\phi }=0\text{.}  \label{eq.24}
\end{equation}

\subsubsection{\protect\bigskip The SCC Conformal Einstein Frame}

If the following conditions hold

i)\qquad $\widetilde{\Box }\ln \Omega =0$, ii) $\qquad \omega =-\frac{3}{2}$
and iii)$\qquad \nabla _{\alpha }\widetilde{\phi }=0$, then the EF
Langrangian density equation \ref{eq.13} reduces to 
\begin{equation}
L^{SCC}[\widetilde{g}]=\frac{\sqrt{-\widetilde{g}}}{16\pi G_{N}}\widetilde{R}%
+\widetilde{L}_{matter}^{SCC}[\widetilde{g}]\text{ ,}  \label{eq.25}
\end{equation}%
where matter is now minimally connected and the conformal transformation of
the Lagrangian density reduces to that of canonical GR.

The energy-momentum tensor of matter is thereby defined by 
\begin{equation}
\widetilde{T}_{M\mu \nu }=\frac{2}{\sqrt{-\widetilde{g}}}\frac{\partial }{%
\partial \widetilde{g}^{\mu \nu }}\left( \sqrt{-\widetilde{g}}\widetilde{L}%
_{matter}\right) \text{ .}  \label{eq.25a}
\end{equation}

In SCC do these three conditions actually hold to achieve this vast
simplification?

The first holds as a static vacuum condition because, under the SCC
conformal factor $\Omega $, the term $\widetilde{\Box }\ln \Omega $ becomes $%
\widetilde{\Box }\widetilde{\Phi }_{N}\left( \widetilde{x}^{\mu }\right) $
and in a harmonic coordinate system this reduces to $\widetilde{\nabla }^{2}%
\widetilde{\Phi }_{N}\left( \widetilde{x}^{\mu }\right) $, which equals zero 
\textit{in vacuo.}

The second condition, $\omega =-\frac{3}{2}$, will shortly be shown to be
the case.

The third holds when the EF is defined as that in which $\tilde{\phi}$ is
constant.

Therefore, for a static metric \textit{in vacuo,} such as in the
Schwarzschild solution, SCC reduces to GR. As experiments verifying GR have
only tested this situation [$R_{\mu \nu }=0$], they will also concur with
SCC. Other tests in which condition i) does not hold will be able to resolve
the degeneracy between the two theories.

\subsection{ Determining the 'creation' equation \protect\ref{eq.6}}

In order to determine $T_{\phi \,\mu \nu }$ and $f_{\nu }\left( \phi \right) 
$ in equations \ref{eq.4} and \ref{eq.6} equation \ref{eq.4} is written in
the following mixed tensor form%
\begin{equation}
T_{M\;\nu }^{\;\mu }=\frac{\phi }{8\pi }\left( R_{\;\nu }^{\mu }-\frac{1}{2}%
\delta _{\;\nu }^{\mu }R\right) -T_{\phi \;\nu }^{\;\mu }  \label{eq.26}
\end{equation}%
and Weinberg's method is followed, (Weinberg, 1972, pages 158-160, equations
7.3.4-7.3.12). The most general form of $T_{\phi \;\nu }^{\;\mu }$ using two
derivatives of one or two $\phi $ fields and $\phi $ itself is 
\begin{eqnarray}
T_{\phi \;\nu }^{\;\mu } &=&A(\phi )g^{\mu \sigma }\nabla _{\sigma }\phi
\nabla _{\nu }\phi +B(\phi )\delta _{\;\nu }^{\mu }\nabla _{\sigma }\phi
g^{\rho \sigma }\nabla _{\rho }\phi   \label{eq.27} \\
&&+C(\phi )g^{\mu \sigma }\nabla _{\sigma }\nabla _{\nu }\phi +D(\phi
)\delta _{\;\nu }^{\mu }\Box \phi 
\end{eqnarray}%
and covariantly differentiating this produces%
\begin{eqnarray}
\nabla _{\mu }T_{\phi \;\nu }^{\;\mu } &=&\left[ A^{\prime }(\phi
)+B^{\prime }(\phi )\right] g^{\mu \nu }\nabla _{\nu }\phi \nabla _{\nu
}\phi \nabla _{\mu }\phi   \label{eq.27a} \\
&&+\left[ A(\phi )+D^{\prime }(\phi )\right] \nabla _{\nu }\phi \Box \phi  \\
&&+\left[ A(\phi )+2B(\phi )+C^{\prime }(\phi )\right] g^{\mu \sigma }\nabla
_{\sigma }\nabla _{\nu }\phi \nabla _{\mu }\phi   \notag \\
&&+D(\phi )\nabla _{\nu }\left( \Box \phi \right) +C(\phi )\Box (\nabla
_{\nu }\phi )\text{ ,}  \notag
\end{eqnarray}

where a prime ($^{\prime }$) is differentiation w.r.t $\phi $. In order to
examine the violation of the equivalence principle use is made of the
Bianchi identities and the identity (observing the sign convention) 
\begin{equation}
\nabla _{\sigma }\phi R_{\;\nu }^{\sigma }=\nabla _{\nu }\left( \Box \phi
\right) -\Box \left( \nabla _{\nu }\phi \right) \text{.}  \label{eq.28}
\end{equation}%
Covariantly differentiating equation \ref{eq.26}, and remembering the
Bianchi identities, yields 
\begin{equation}
\nabla _{\mu }T_{M\;\nu }^{\;\mu }=\frac{\nabla _{\mu }\phi }{8\pi }\left(
R_{\;\nu }^{\mu }-\frac{1}{2}\delta _{\;\nu }^{\mu }R\right) -\nabla _{\mu
}T_{\phi \;\nu }^{\;\mu }\text{ .}  \label{eq.29}
\end{equation}%
Now if we take the trace of equation \ref{eq.4} we obtain%
\begin{equation}
R=-\frac{8\pi }{\phi }\left[ T_{M\;\sigma }^{\;\sigma }+T_{\phi \;\sigma
}^{\;\sigma }\right] \text{ ,}  \label{eq.30}
\end{equation}%
with 
\begin{equation}
T_{\phi \;\sigma }^{\;\sigma }=\left[ A(\phi )+4B(\phi )\right] g^{\sigma
\rho }\nabla _{\rho }\phi \nabla _{\sigma }\phi +\left[ C(\phi )+4D(\phi )%
\right] \Box \phi \text{ ,}  \label{eq.31}
\end{equation}%
and from equation \ref{eq.1} 
\begin{equation}
T_{M\;\sigma }^{\;\sigma }=\frac{1}{4\pi \lambda }\Box \phi \text{ .}
\label{eq.32}
\end{equation}%
Substituting equations \ref{eq.31} and \ref{eq.32} in equation \ref{eq.30}
yields 
\begin{equation}
R=-\frac{8\pi }{\phi }\left\{ \left[ A(\phi )+4B(\phi )\right] g^{\sigma
\rho }\nabla _{\rho }\phi \nabla _{\sigma }\phi +\left[ C(\phi )+4D(\phi )+%
\frac{1}{4\pi \lambda }\right] \Box \phi \right\} \text{ .}  \label{eq.33}
\end{equation}%
While equations \ref{eq.27a}, \ref{eq.28} and \ref{eq.33} substituted in \ref%
{eq.29} produce 
\begin{eqnarray}
\nabla _{\mu }T_{M\;\nu }^{\;\mu } &=&-\frac{1}{8\pi }\nabla _{\nu }(\Box
\phi )+\frac{1}{8\pi }\Box (\nabla _{\nu }\phi )+\frac{1}{2\phi }\left[
A(\phi )+4B(\phi )\right] g^{\mu \sigma }\nabla _{\sigma }\phi \nabla _{\mu
}\phi \nabla _{\nu }\phi   \label{eq.34} \\
&&\ +\frac{1}{2\phi }\left[ C(\phi )+4D(\phi )+\frac{1}{4\pi \lambda }\right]
\nabla _{\nu }\phi \Box \phi   \notag \\
&&\ -\left[ A^{\prime }(\phi )+B^{\prime }(\phi )\right] g^{\mu \sigma
}\nabla _{\sigma }\phi \nabla _{\nu }\phi \nabla _{\mu }\phi -\left[ A(\phi
)+D^{\prime }(\phi )\right] \nabla _{\nu }\phi \Box \phi   \notag \\
&&\ -\left[ A(\phi )+2B(\phi )+C^{\prime }(\phi )\right] g^{\mu \sigma
}\nabla _{\sigma }\nabla _{\nu }\phi \nabla _{\mu }\phi   \notag \\
&&-D(\phi )\nabla _{\nu }\left( \Box \phi \right) -C(\phi )\Box (\nabla
_{\nu }\phi )\text{ .}
\end{eqnarray}%
If the Principle of Mutual Interaction, equation \ref{eq.6}, is applied 
\begin{equation*}
\nabla _{\mu }T_{M\quad \nu }^{.\quad \mu }=f_{\nu }\left( \phi \right) \Box
\phi 
\end{equation*}%
So the coefficients of:$(\Box \phi );_{\nu }$; $\Box (\phi ;_{\nu })$, $\phi
;^{\mu };_{\nu }\phi ;_{\mu }$, and $\phi ;^{\mu }\phi ;_{\mu }\phi ;_{\nu }$%
, must vanish in equation \ref{eq.34}, but those of $\phi ;_{\nu }\Box \phi $
must satisfy equation \ref{eq.6}. This yields five equations to solve for
the five functions; $A(\phi )$, $B(\phi )$, $C(\phi )$, $D(\phi )$ and $%
f_{\nu }(\phi )$. 
\begin{equation*}
\begin{tabular}{llllll}
\underline{Term} &  & \underline{Coefficients ( = 0)} &  & \underline{%
Solution} &  \\ 
$(\Box \phi );_{\nu }$ &  & $-\frac{1}{8\pi }-D(\phi )=0$ &  & $D(\phi )=-%
\frac{1}{8\pi }$ & (i) \\ 
$\Box (\phi ;_{\nu })$ &  & $+\frac{1}{8\pi }-C(\phi )=0$ &  & $C(\phi )=+%
\frac{1}{8\pi }$ & (ii) \\ 
$\phi ;^{\mu };_{\nu }\phi ;_{\mu }$ &  & $A(\phi )+2B(\phi )+C^{\prime
}(\phi )=0$ &  & $A(\phi )=-2B(\phi )$ & (iii) \\ 
$\phi ;^{\mu }\phi ;_{\mu }\phi ;_{\nu }$ &  & $\frac{1}{2\phi }\left[
A(\phi )+4B(\phi )\right] $ &  &  &  \\ 
&  & $-\left[ A^{\prime }(\phi )+4B^{\prime }(\phi )\right] =0$ &  &  & (iv)%
\end{tabular}%
\ 
\end{equation*}%
Substituting equation (iii) into (iv) 
\begin{equation}
\frac{B^{\prime }(\phi )}{B(\phi )}=-\frac{1}{\phi }\text{,}  \label{eq.35}
\end{equation}%
which has the solution 
\begin{equation}
B(\phi )=\frac{k}{\phi }\text{ ,}  \label{eq.36}
\end{equation}%
where $k$ is a constant, and therefore by equation (iii) 
\begin{equation}
A(\phi )=-\frac{2k}{\phi }\text{ .}  \label{eq.37}
\end{equation}%
If $\kappa $ is now written as 
\begin{equation*}
k=-\frac{\omega }{16\pi }
\end{equation*}%
the non-unique solution is obtained 
\begin{equation}
\begin{tabular}{llll}
& $A(\phi )=\frac{\omega }{8\pi \phi }$ &  & $B(\phi )=-\frac{\omega }{16\pi
\phi }$ \\ 
& $C(\phi )=\frac{1}{8\pi }$ &  & $D(\phi )=-\frac{1}{8\pi }$%
\end{tabular}%
\ \text{ .}  \label{eq.38}
\end{equation}

This solution looks the same as the BD solution except that $\omega $ is as
yet undetermined. In BD there is a fifth redundant equation, which may be
used here in SCC to determine $\omega $. A solution for $\nabla _{\mu
}T_{M\;\nu }^{\;\mu }$ is obtained by substituting equation \ref{eq.38} into
equation \ref{eq.34} and examining the coefficients of $\nabla _{\nu }\phi
\Box \phi $. This results in a fifth equation that determines $\omega $: 
\begin{equation}
\nabla _{\mu }T_{M\quad \nu }^{.\quad \mu }=\left( \frac{1}{16\pi \phi }-%
\frac{1}{4\pi \phi }+\frac{1}{8\pi \lambda \phi }-\frac{\omega }{8\pi \phi }%
\right) \nabla _{\nu }\phi \Box \phi \text{,}  \label{eq.39}
\end{equation}%
which can be written as 
\begin{equation}
\nabla _{\mu }T_{M\quad \nu }^{.\quad \mu }=\frac{\kappa }{8\pi }\frac{%
\nabla _{\nu }\phi }{\phi }\Box \phi \text{ ,}  \label{eq.40}
\end{equation}%
so from equation \ref{eq.6}%
\begin{equation}
f_{\nu }\left( \phi \right) =\frac{\kappa }{8\pi }\frac{\nabla _{\nu }\phi }{%
\phi }  \label{eq.41}
\end{equation}%
where 
\begin{equation}
\kappa =\frac{1}{\lambda }-\frac{3}{2}-\omega \text{ .}  \label{eq.42}
\end{equation}%
$\kappa $ can be thought of as an undetermined \textquotedblright creation
coefficient\textquotedblright . Note however that if $\kappa =0$ then 
\begin{equation}
\omega =\frac{1}{\lambda }-\frac{3}{2}  \label{eq.43}
\end{equation}%
and the BD field equations have been recovered as to be expected. however,
in general, if $\kappa \neq 0$ then we have a modified version of BD. In
particular if the condition 
\begin{equation}
\kappa =\frac{1}{\lambda }  \label{eq.44}
\end{equation}%
holds then we will have 
\begin{equation}
\omega =-\frac{3}{2}  \label{eq.45}
\end{equation}%
without $\lambda \rightarrow \infty $. It will be shown that applying the
PMI determines both $\kappa $ and $\lambda $\ to be unity and therefore
condition equation \ref{eq.44} actually does hold. Furthermore with these
values for $\lambda $ and $\kappa $ equation \ref{eq.40} becomes 
\begin{equation}
\nabla _{\mu }T_{M\nu }^{\;\mu }=\frac{1}{2}\frac{1}{\phi }\nabla _{\nu
}\phi T_{M}\text{ ,}  \label{eq.67}
\end{equation}%
which is the same dynamical equation as in the JF of GR. (Quiros, 2001
equation 3.3)

Furthermore when $\nabla _{\mu }\phi =0$ equation \ref{eq.67} reduces to
equation \ref{eq.3} and in this immediate locality where $g_{\mu \nu
}\rightarrow \eta _{\mu \nu }$ and $\phi =$ $\phi _{c}$, a minima, SCC
reduces to SR and the theory admits a ground state solution.

As with any scalar-tensor theory the question must be asked: "Which is the
physical frame?" In SCC not only is the equivalence principle preserved in
the EF, but also, if $\omega =-\frac{3}{2}$, the scalar field energy density
is non-negative in the JF (see analysis by Santiago \& Silbergleit, 2000).
Therefore both EF and JF can be considered to be physical frames for the
theory and which is appropriate for any particular situation depends on the
choice of the invariant standard of measurement.

\subsection{The choice of conformal frame}

In addressing the question as to which conformal frame is appropriate it
must be noted that a further form of the equations must be considered. There
has been discussion in the literature about the correct form of the BD field
equation motivated by the scalar field energy density not being positive
definite in BD. This problem does not arise when $\omega =-\frac{3}{2}$ as
in SCC, (Santiago \& Silbergleit, 2000), however the subsequent analysis is
pertinent to this theory. In the standard formulation of BD the
energy-momentum tensor of the scalar field contains the second derivatives
of $\phi $ , which are necessarily convoluted with the gravitational terms
of the affine connection. The \textquotedblright
corrected\textquotedblright\ version is formulated so that this is not the
case and this version is called the \textquotedblright '\textbf{true}'
stress-energy tensor\textquotedblright\ while the original version is called
the \textquotedblright '\textbf{effective}' stress-energy
tensor.\textquotedblright\ (Quiros, 2001) Whether the need for this
correction exists in the $\omega =-\frac{3}{2}$ case is obscure but cast
into its 'corrected' form the left hand side of the gravitational field
equation, the Einstein tensor $G_{\mu \nu }$ becomes the 'affine' Einstein
tensor $^{\gamma }G_{\mu \nu }$ and in this case the whole JF equation
becomes 
\begin{equation}
^{\gamma }G_{\mu \nu }=\frac{8\pi }{\phi }T_{M\mu \nu }+\frac{\left( \omega +%
\frac{3}{2}\right) }{\phi ^{2}}\left( \nabla _{\mu }\phi \nabla _{\nu }\phi -%
\frac{1}{2}g_{\mu \nu }g^{\alpha \beta }\nabla _{\alpha }\phi \nabla _{\beta
}\phi \right)   \label{eq.67a}
\end{equation}%
so in the SCC3 case with $\omega =-\frac{3}{2}$ this reduces down to 
\begin{equation}
^{\gamma }G_{\mu \nu }=\frac{8\pi }{\phi }T_{M\mu \nu }\text{ .}
\label{eq.67b}
\end{equation}%
In other words in this representation of the theory the field equation
reduces to that of the original SCC2, equation \ref{eq.5}, with its
attendant problems of the definition of the metric. In an earlier paper
(Barber, 2002) the original, 'effective' scalar field energy-momentum tensor
was treated as the physical representation of the theory in the JF and used
to make the prediction for the Gravity Probe B geodetic precession. It
proved to be false. Here the effective stress energy tensor shall be used
consistently to determine the value of $\omega $\ and explore the action of
the scalar field force but predictions of the theory concerning the motion
of massive test-particles must be calculated using the \textbf{true}
stress-energy tensor in which case the field equations \textit{in vacuo}
reduce to that of GR ($R_{\mu \nu }=0=\allowbreak 0$). For example the
static Schwarzschild solution is identical to GR and the Robertson
parameters, which define the metric experienced in gravitational
experiments, are therefore:%
\begin{equation}
\alpha _{true}=1\text{, \ }\beta _{true}=1\text{ \ and \ }\gamma _{true}=1
\label{eq.67c}
\end{equation}

Using these values the theory is consistent with all the standard tests
including the geodetic and Lens Thirring precession measurements of the
Gravity Probe B experiment.

The GP-B experiment has demonstrated that (contrary to the paper by this
author, 2002) the 'true' stress energy tensor, hereafter described as the
'true JF', must be used in predictions of the behaviour of massive
particles. However massless particles such as photons, do not travel on
geodesics of the metric in the true JF (as noted by Brans, 1987), therefore
the path of light has to be evaluated using the \textbf{effective}
stress-energy scalar field energy-momentum tensor in the 'effective JF'.
Furthermore, as the scalar field is disconnected from the gravitational
field and thereby becomes a 'ghost' field in the true JF as well as in the
EF, then the effective JF has to be used when calculating scalar and
gravitational fields. The EF can be used when 'atomic' units (with time
measured by an atomic clock) have been chosen to interpret observations.

There are two questions to ask in order that a Weyl metric may be set up
spanning extended space-time; \textquotedblright What is the invariant
standard by which objects are to be measured?\textquotedblright\ and
\textquotedblright How is that standard to be transmitted from event to
event in order that the comparison can be made?\textquotedblright\ In both
GR and the EF\ of SCC the invariant is defined by the principle of the
conservation of energy-momentum to be invariant 'rest' mass. The standard of
measurement therefore becomes that of 'fixed' (atomic i.e. steel or
platinum) rulers and 'regular' atomic clocks. In the JF of SCC, on the other
hand, the invariant is defined by the principle of the local conservation of
energy, and therefore the standard of measurement is defined to be that of a
carefully defined 'standard photon', with its frequency (inverse)
determining the standard of time and space measurement, and its energy
determining the standard of mass, all defined in the CoM, Machian, frame of
reference. The concept of a standard photon is refined as the theory
unfolds, however note that cosmological time in the JF is defined by the
radiation of the CMB.

\subsection{The field equations}

The form of the scalar field is now developed in the effective JF in which
the complete set of field equations are now:

1.\ The scalar field equation \ref{eq.1}.%
\begin{equation*}
\Box \phi =4\pi \lambda T_{M}^{\;}\text{ }
\end{equation*}

2.\ The gravitational field equation\ \ref{eq.12a}%
\begin{eqnarray*}
R_{\mu \nu }-\frac{1}{2}g_{\mu \nu }R &=&\frac{8\pi }{\phi }T_{M\mu \nu }+%
\frac{\omega }{\phi ^{2}}\left( \nabla _{\mu }\phi \nabla _{\nu }\phi -\frac{%
1}{2}g_{\mu \nu }\nabla _{\sigma }\phi g^{\sigma \rho }\nabla _{\rho }\phi
\right) \\
&&+\frac{1}{\phi }\left( \nabla _{\mu }\nabla _{\nu }\phi -g_{\mu \nu }\Box
\phi \right) \text{ ,}
\end{eqnarray*}%
where $\omega =\frac{1}{\lambda }-\frac{3}{2}-\kappa $ is a constant and $%
\lambda $ and $\kappa $ are coefficients yet to be determined.

3.\ The \textquotedblright creation\textquotedblright\ field equation \ref%
{eq.40}, which together with equation \ref{eq.1} becomes,

\begin{equation}
\nabla _{\mu }T_{M\nu }^{\;\mu }=\frac{\kappa }{8\pi }\frac{\nabla _{\nu
}\phi }{\phi }\Box \phi =\frac{\kappa \lambda }{2}\frac{\nabla _{\nu }\phi }{%
\phi }T_{M}\text{.}  \label{eq.68}
\end{equation}

\section{Determining $\protect\lambda $ and $\protect\kappa $ to derive $%
\protect\omega $}

The source of curvature $S_{\mu \nu }$ is defined by 
\begin{equation}
R_{\mu \nu }=\frac{8\pi }{\phi }S_{\mu \nu }  \label{eq.69}
\end{equation}%
where $S_{\mu \nu }$ is derived from equation \ref{eq.12a} to be 
\begin{eqnarray}
S_{\mu \nu } &=&T_{M\,\mu \nu }-\frac{1}{2}\left( 1-\frac{1}{2}\lambda
\right) g_{\mu \nu }T_{M\;\sigma }^{\;\sigma }+\frac{\omega }{8\pi \phi }%
\phi ;_{\mu }\phi ;_{\nu }  \label{eq.70} \\
&&\ +\frac{1}{8\pi }\phi ;_{\mu };_{\nu }\text{ .}  \notag
\end{eqnarray}%
The gravitational field equation can be written 
\begin{eqnarray}
R_{\mu \nu } &=&\frac{8\pi }{\phi }\left[ T_{M\,\mu \nu }-\frac{1}{2}\left(
1-\frac{1}{2}\lambda \right) g_{\mu \nu }T_{M\;\sigma }^{\;\sigma }\right] +%
\frac{\omega }{\phi ^{2}}\phi ;_{\mu }\phi ;_{\nu }  \label{eq.71} \\
&&\ +\frac{1}{\phi }\phi ;_{\mu };_{\nu }  \notag
\end{eqnarray}%
so $R_{\mu \nu }$ can be written in terms of the BD parameter, where $\varpi
=\frac{1}{\lambda }-\frac{3}{2}$ as follows: 
\begin{eqnarray}
R_{\mu \nu } &=&\frac{8\pi }{\phi }\left[ T_{M\,\mu \nu }-\left( \frac{%
1+\varpi }{3+2\varpi }\right) g_{\mu \nu }T_{M\;\sigma }^{\;\sigma }\right]
\label{eq.72} \\
&&\ +\frac{\varpi }{\phi ^{2}}\phi ;_{\mu }\phi ;_{\nu }+\frac{1}{\phi }\phi
;_{\mu };_{\nu }-\frac{\kappa }{\phi ^{2}}\phi ;_{\mu }\phi ;_{\nu }  \notag
\end{eqnarray}%
which is the same as the equivalent equation in the BD theory except with
the addition of the last term which includes the \textquotedblright creation
coefficient\textquotedblright\ $\kappa $. This expression was used in the
2002 paper to enable comparison of the solution with the standard BD theory.

\subsection{The Post-Newtonian Approximation of the One-Body Problem}

The development of the One Body Problem in SCC is the same as in the earlier
paper (Barber, 2002), where a full treatment may be found, so here a summary
is given. The equations in the One Body Problem differ from BD only
insignificantly, as the extra SCC term involving $\kappa $\ is time
dependent and drops out in the static case.

Consider the stationary gravitational and scalar fields around a static,
spherically symmetric, mass embedded in a cosmological space-time. In such
an embedding the value of the scalar $\phi $ defining inertial mass is
assumed to asymptotically approach a \textquotedblright
cosmological\textquotedblright\ value $G_{0}^{\;-1}$ which holds
\textquotedblright at great distance\textquotedblright\ from any large
masses. $\phi $ is determined in the inertial, Lorentz frame of reference of
the Centre of Mass using electromagnetic methods and this is the origin of
our coordinate system. 
\begin{equation}
\phi =G_{0}^{\;-1}\left( 1+\epsilon \right)  \label{eq.73}
\end{equation}%
where $G_{0}$ is a constant of dimension and order $G_{N}$, and $\epsilon $
a scalar field defined by 
\begin{equation}
\Box \epsilon =\epsilon ;^{\sigma };_{\sigma }=\frac{8\pi }{3+2\varpi }%
G_{0}T_{\;\sigma }^{\sigma }  \label{eq.74}
\end{equation}%
in which $\epsilon \rightarrow 0$ as $r\rightarrow \infty $ and we note $%
\varpi $ is the BD parameter $\varpi =\frac{1}{\lambda }-\frac{3}{2}$. $%
T_{\mu \nu }$ is the energy-momentum tensor of ordinary matter and energy
excluding the energy of the $\phi $ field.

The gravitational field equation \ref{eq.72} now becomes 
\begin{eqnarray}
R_{\mu \nu } &=&8\pi G_{0}\left( 1+\epsilon \right) ^{-1}\left[ T_{M\,\mu
\nu }-\left( \frac{1+\varpi }{3+2\varpi }\right) g_{\mu \nu }T_{M\;\sigma
}^{\;\sigma }\right]  \label{eq.75} \\
&&\ +\frac{\varpi }{\left( 1+\epsilon \right) ^{2}}\epsilon ;_{\mu }\epsilon
;_{\nu }+\frac{1}{\left( 1+\epsilon \right) }\epsilon ;_{\mu };_{\nu }-\frac{%
\kappa }{\left( 1+\epsilon \right) ^{2}}\epsilon ;_{\mu }\epsilon ;_{\nu }%
\text{ ,}  \notag
\end{eqnarray}%
which again is the same as the equivalent BD equation except with the
addition of the last term which includes $\kappa $.

It is this last term that drops out when we consider the stationary fields
of the Post-Newtonian Approximation (PNA) and in which only slowly moving
particles are considered. If $\overline{r}$ and $\overline{v}$ are typical
distances and velocities of the system then the components of the metric and
the Ricci tensor are expressed in powers of the parameters $\frac{GM}{%
\overline{r}}$ and $\overline{v}^{2}$ and the PNA requires an expansion of
these parameters to one order beyond Newtonian mechanics.

From equation \ref{eq.74} $\epsilon $ has the expansion 
\begin{equation}
\epsilon =\overset{2}{\,\epsilon }+\overset{4}{\epsilon }+...  \label{eq.76}
\end{equation}%
where $\overset{N}{\epsilon }$ is of the order $\overline{v}^{N}$ and in
particular 
\begin{equation}
\nabla ^{2}\overset{2}{\epsilon }\,=-\frac{8\pi G_{0}}{3+2\varpi }\overset{0}%
{T}^{00}  \label{eq.77}
\end{equation}%
Substituting the PNA formulas into the Ricci tensor and applying the result
to the field equation \ref{eq.74} we obtain 
\begin{equation}
\nabla ^{2}\overset{2}{g}_{00}=-8\pi G_{0}\left( \frac{2\varpi +4}{2\varpi +3%
}\right) \overset{0}{T}^{00}\text{ ,}  \label{eq.78}
\end{equation}%
From this equation it follows that the usual relation between $\overset{2}{g}%
_{00}$and the purely gravitational Newtonian potential $\Phi _{m}$ holds by
defining $\Phi _{m}$ as 
\begin{equation}
\nabla ^{2}\Phi _{m}=4\pi G_{m}\overset{0}{T}^{00}  \label{eq.79}
\end{equation}%
so normalized that 
\begin{equation*}
\Phi _{m}\left( \infty \right) =0
\end{equation*}%
where $G_{m}$ is the metric gravitational \textquotedblright
constant\textquotedblright\ associated with the curvature of space-time
measured in the limit $r\rightarrow \infty $. Then $G_{m}$ is related to $%
G_{0}$ by the relationship 
\begin{equation}
G_{m}=\left( \frac{2\varpi +4}{2\varpi +3}\right) G_{0}  \label{eq.80}
\end{equation}%
Consequently it is important to note that in the BD theory, where $%
G_{m}\equiv G_{N}$, the definitions of $G_{0}$, $\phi $, and $\psi $ in
equations \ref{eq.2} and \ref{eq.73} give the result 
\begin{equation}
\psi =\left( \frac{2\varpi +4}{2\varpi +3}\right)  \label{eq.81}
\end{equation}%
Thus $\psi $ is not necessarily unity and has to be determined in this
calculation. The relationship between $\epsilon $ and $\Phi _{m}$ is derived
from the equations \ref{eq.77} and \ref{eq.79} to be 
\begin{equation}
\overset{2}{\epsilon }\,=-\frac{1}{\varpi +2}\Phi _{m}  \label{eq.82}
\end{equation}%
Note the solution for $\Phi _{m}$ around a spherically symmetric mass in
vacuo is given by 
\begin{equation*}
\Phi _{m}=-\frac{G_{m}M}{r}
\end{equation*}%
therefore 
\begin{equation}
\overset{2}{\epsilon }\,=+\frac{G_{m}M}{\left( 2+\varpi \right) r}
\label{eq.83}
\end{equation}%
In the static PNA solution there is no difference between this theory and
BD. Hence, as with BD, the gravitational field outside a static, spherically
symmetric mass depends on $M$ alone but not any other property of the mass.
Also the \textit{effective} (but not the \textit{true}) Robertson parameters
for this theory are also given by the same formulas as in BD 
\begin{equation}
\alpha _{eff}=1\text{ , }\beta _{eff}=1\text{ , }\gamma _{eff}=\frac{\varpi
+1}{\varpi +2}=\frac{2-\lambda }{2+\lambda }\text{.}  \label{eq.84}
\end{equation}

\subsection{At the centre of mass}

The violation of the equivalence principle in the JF introduces two
potentially serious complications; the definition of the metric, and the
provision of a suitable frame of reference that admits a ground state
solution in which consistent measurements of proper time, and therefore
distance, can be made. It will be shown that both these complications are
resolved by the PMI.

It has been shown that under the PMI photons travel along null-geodesics in
the SCC JF. Another way of stating this is to note that conformal
transformations from the EF canonical GR do not affect the trajectories of
trace-free entities. Hence there is no problem defining the metric.

In order to examine the second complication consider the origin of our
coordinate system in the static, spherically symmetric, case, which is the
centre of mass of the system. In Relativity theory the centroid of an
isolated system with energy-momentum tensor $T^{\mu \nu }$ and total
4-momentum $P^{\alpha }$, when observed by an observer\textit{\ O }with a
4-velocity $U^{\alpha }$ at his Lorentz time $x^{0}=t$ and in his own
Lorentz frame, is defined by 
\begin{equation}
X_{\;\mu }^{j}\left( t\right) =\left( \frac{1}{P^{0}}\right) \underset{%
x_{0}=t}{\int }x^{j}T^{00}d^{3}x  \label{eq.85}
\end{equation}%
and the co-moving centroid associated with the rest frame of the system is
defined to be its Centre of Mass (CoM). At the CoM the resultant of all
gravitational forces vanishes hence so does $\underline{\nabla }\Phi _{N}$.
Furthermore $\phi =\phi \left( \Phi _{N\text{ }}\right) $, therefore with $%
\underline{\nabla }\Phi _{N}=0$ at the CoM, 
\begin{equation}
\text{ }\nabla \phi =0\text{ .}  \label{eq.86}
\end{equation}%
As $\phi \left( x_{\nu }\right) $ is static and not dependent on time we
have for all four $\nu $%
\begin{equation}
\nabla _{\nu }\phi =0\text{ ,}  \label{eq.87}
\end{equation}%
thus at the CoM, by Equation \ref{eq.40}, the PMI yields 
\begin{equation}
\nabla _{\mu }T_{M\;\nu }^{\;\mu }=\frac{\kappa }{8\pi }\frac{1}{\phi }%
\nabla _{\nu }\phi \Box \phi =0\text{ .}  \label{eq.88}
\end{equation}

Hence at the unique location of the centre of mass of the system the
energy-momentum tensor of matter is conserved with respect to covariant
differentiation. Here the theory admits a ground state solution, $g_{\mu \nu
}\rightarrow \eta _{\mu \nu }$ and $\nabla _{\mu }\phi =0$ , here the
equivalence principle holds, even for a massive particle, and here a free
falling physical clock, remaining at rest, records proper time. Distances
can be measured by timing the echo of light rays (radar) using that clock
and the metric properly defined. Also Special Relativity is recovered as
here the metric is Minkowskian and standards of mass, length, time and the
physical constants defined for atoms, together with potential energy, retain
their classical meaning. Such a standard defined atom emits a 'reference'
photon mentioned earlier, which in the JF is transmitted across space-time
with invariant energy and frequency.

There may be objections to the acceptance of a preferred frame of reference
because this means giving up the principle of relativity. Note, however,
that even in GR the CoM is a unique, \textquotedblright
preferred\textquotedblright , frame in that although all inertial frames are
equivalent as far as the conservation of the four momentum vector, $P^{\mu }$%
, is concerned, it is the only frame of reference in which the total
mass-energy, $P^{0}$, of the system is also conserved over time. In SCC the
CoM preferred frame of reference may be selected if and only if energy is to
be locally conserved, otherwise the equations are manifestly covariant.

In order that the theory may be fully determined and consequently tested, it
is necessary to evaluate $\lambda $, $\kappa $ and $\varphi $. This is
possible by requiring consistency in the above stated 'preferred' frame of
reference, the frame which is 'selected' by Mach's Principle, that of the
CoM of the system.

\subsection{\protect\bigskip Solving for $\protect\phi $}

The effect of allowing $T_{M\;\nu ;\mu }^{\;\mu }\neq 0$ in the effective JF
has now to be calculated and its effect included in the modelling of
experiments of slowly moving particles. That is, the violation of the
equivalence principle will produce a force $G_{\nu }$ that will perturb
particles, but not photons, from their geodesic world lines. The force
density is given by 
\begin{equation}
G_{\nu }=T_{M\;\nu ;\mu }^{\;\mu }\text{ .}  \label{eq.89}
\end{equation}%
In order to calculate this effect $\phi $ has to be determined to the third
order of accuracy, ($\overset{4}{\epsilon }$), and this is possible both in
BD and SCC. In the PNA solution to the One-Body problem the solution for $%
\phi $ obtained from equations \ref{eq.73}, \ref{eq.76} and \ref{eq.83} is 
\begin{equation}
\phi =G_{0}^{-1}\left[ 1+\frac{G_{m}M}{\left( 2+\varpi \right) r}+...\right]
\label{eq.90}
\end{equation}%
and when the metric takes the standard form of the Robertson expansion 
\begin{eqnarray}
d\tau ^{2} &=&\left( 1-\frac{2G_{m}M}{r}+...\right) dt^{2}-\left( 1+\frac{%
2\gamma G_{m}M}{r}+...\right) dr^{2}  \label{eq.91} \\
&&\ \ +r^{2}d\theta ^{2}+r^{2}\sin ^{2}\theta d\varphi ^{2}\text{ ,}  \notag
\end{eqnarray}%
where $\gamma =\frac{\left( \varpi +1\right) }{\left( \varpi +2\right) }$ .
Then, as $\Box ^{2}\phi =0$ in vacuo, 
\begin{equation}
\Box ^{2}\phi =\frac{d^{2}\phi }{dx^{\lambda \,\,2}}+\Gamma _{\mu \lambda
}^{\mu }\frac{d\phi }{dx^{\lambda \,}}=0\text{ ,}  \label{eq.92}
\end{equation}%
where the affine connection $\Gamma _{\mu \lambda }^{\mu }$ is given by 
\begin{equation}
\Gamma _{\mu \nu }^{\lambda }=\frac{1}{2}g^{\lambda \rho }\left( \frac{%
\partial g_{\rho \mu }}{\partial x^{\nu }}+\frac{\partial g_{\rho \nu }}{%
\partial x^{\mu }}-\frac{\partial g_{\mu \nu }}{\partial x^{\rho }}\right) 
\text{ .}  \label{eq.93}
\end{equation}%
As $g_{\mu \nu }$ is diagonal and $\phi =\phi \left( r\right) $, the only
non-vanishing components of the affine connection are 
\begin{eqnarray}
\Gamma _{tt}^{t} &=&\frac{1}{2}g^{00}\frac{dg_{00}}{dr}=+\frac{G_{m}M}{r^{2}}%
\left[ 1+O\left( \frac{G_{m}M}{r}\right) ...\right] \text{ ,}  \label{eq.94}
\\
\Gamma _{rr}^{r} &=&\frac{1}{2}g^{rr}\frac{dg_{rr}}{dr}=-\frac{\left( \varpi
+1\right) }{\left( \varpi +2\right) }\frac{G_{m}M}{r^{2}}\left[ 1+O\left( 
\frac{G_{m}M}{r}\right) ...\right] \text{ ,}  \notag \\
\Gamma _{\phi r}^{\phi } &=&\frac{1}{2}g^{\phi \phi }\frac{dg_{\phi \phi }}{%
dr}=\frac{1}{r}\text{, and\qquad }\Gamma _{\theta r}^{\theta }=\frac{1}{2}%
g^{\theta \theta }\frac{dg_{\theta \theta }}{dr}=\frac{1}{r}\text{ ,}  \notag
\end{eqnarray}%
and equation \ref{eq.92} becomes 
\begin{equation}
\Box ^{2}\phi =\frac{d^{2}\phi }{dr^{\,2}}+\left\{ \left[ 1-\frac{\left(
\varpi +1\right) }{\left( \varpi +2\right) }\right] \frac{G_{m}M}{r^{2}}+%
\frac{2}{r}+...\right\} \frac{d\phi }{dr}=0\text{ .}  \label{eq.95}
\end{equation}%
Integrating twice w.r.t. $r$, and expanding the exponential with $\frac{%
G_{m}M}{r}\ll 1$ , produces a solution with two integration constants, $%
k_{1} $ and $k_{2}$; 
\begin{equation}
\phi =k_{1}+\frac{k_{2}}{r}+\frac{1}{2\left( \varpi +2\right) }\frac{G_{m}M}{%
r^{2}}+...\text{ .}  \label{eq.96}
\end{equation}%
Comparing coefficients with equation \ref{eq.90} evaluates $k_{1}$ and $%
k_{2} $ so 
\begin{equation}
\phi =G_{0}^{-1}\left\{ 1+\frac{G_{m}M}{\left( 2+\varpi \right) r}+\frac{1}{2%
}\left[ \frac{G_{m}M}{\left( 2+\varpi \right) r}\right] ^{2}+...\right\} 
\text{ ,}  \label{eq.97}
\end{equation}%
and therefore, to the accuracy of the post-post Newtonian approximation, 
\begin{equation}
\phi =\phi _{0}\exp \left[ \frac{G_{m}M}{\left( 2+\varpi \right) r}\right] 
\text{ .}  \label{eq.98}
\end{equation}%
Therefore 
\begin{equation}
\frac{1}{\phi }\frac{d\phi }{dr}=-\frac{G_{m}M}{\left( 2+\varpi \right) r^{2}%
}\text{ .}  \label{eq.99}
\end{equation}

\subsection{The scalar field acceleration}

The expression for $T_{M\;\nu ;\mu }^{\;\mu }$ for a system of n particles
of rest mass $m_{n}$ is given by 
\begin{equation}
T_{M\;\nu ;\mu }^{\;\mu }=G_{\nu }=\underset{n}{\sum }\delta ^{3}\left\{
x-x_{n}\left( t\right) \right\} g_{\nu \alpha }\frac{d\tau }{dt}\frac{d}{dt}%
\left[ m_{n}\frac{dx^{\alpha }}{d\tau }\right] \text{ ,}  \label{eq.100}
\end{equation}%
where $\delta ^{3}\left\{ x-x_{n}\left( t\right) \right\} $ is the Dirac
delta function, $d\tau $ the proper time defined by 
\begin{equation}
d\tau ^{2}=-g_{\mu \nu }dx^{\mu }dx^{\nu }  \label{eq.101}
\end{equation}%
and $G_{\nu }$ is the force density.

Over the elemental volume of an individual test particle with density of
inertial rest mass $\rho \left( r,t\right) $ becomes 
\begin{equation*}
T_{M\;\nu ;\mu }^{\;\mu }=g_{\alpha \nu }\frac{d\tau }{dt}\frac{d}{dt}\left[
\rho \frac{dx^{\alpha }}{d\tau }\right] \text{ ,}
\end{equation*}%
that is, 
\begin{equation}
T_{M\;\nu ;\mu }^{\;\mu }=g_{\alpha \nu }\left[ \frac{d\rho }{dt}\frac{%
dx^{\alpha }}{dt}+\rho \frac{d^{2}x^{\alpha }}{dt^{2}}\right] \text{ .}
\label{eq.102}
\end{equation}%
But using equation \ref{eq.68} together with the equation of state for a
perfect fluid 
\begin{equation}
T_{M\;\sigma }^{\;\sigma }=3p-\rho =-\rho  \label{eq.103}
\end{equation}%
in the rest frame when the pressure is negligible, we obtain the PMI
solution for the non-conservation of the energy-momentum tensor, 
\begin{equation}
T_{M\;\nu ;\mu }^{\;\mu }=-\frac{\kappa \lambda }{2}\frac{\phi ;_{\nu }}{%
\phi }\rho \text{ .}  \label{eq.104}
\end{equation}%
Therefore the effect of the scalar field force is given by 
\begin{equation}
g_{\alpha \nu }\left[ \frac{d\rho }{dt}\frac{dx^{\alpha }}{dt}+\rho \frac{%
d^{2}x^{\alpha }}{dt^{2}}\right] =-\frac{\kappa \lambda }{2}\frac{\phi
;_{\nu }}{\phi }\rho  \label{eq.105}
\end{equation}%
Now consider the effect of this force on a mass particle momentarily at rest
in the frame of reference of the Centre of Mass, that is: $\frac{dx^{\alpha }%
}{dt}=0$, equation \ref{eq.105} becomes 
\begin{equation}
\frac{d^{2}x^{\alpha }}{dt^{2}}=-g^{\alpha \nu }\frac{\kappa \lambda }{2}%
\frac{\phi ;_{\nu }}{\phi }\text{ .}  \label{eq.106}
\end{equation}

\subsection{Equations of motion}

It is now possible to examine the equations of motion in this theory. At
every space-time event in an arbitrary gravitational field we can specify a
set of coordinates $\xi ^{i}$ in which the local description of space-time
is Minkowskian, with a Special Relativity metric $\eta _{\alpha \beta }$ and
in which a photon has an equation of motion 
\begin{equation}
\frac{d^{2}\xi ^{\alpha }}{d\sigma ^{2}}=0\text{ ,}  \label{eq.107}
\end{equation}%
\begin{equation}
0=-\eta _{\alpha \beta }\frac{d\xi ^{\alpha }}{d\sigma }\frac{d\xi ^{\beta }%
}{d\sigma }\text{ ,}  \label{eq.108}
\end{equation}%
where $\sigma \equiv \xi ^{0}$ is a suitable parameter describing the
null-geodesic. We now consider the equation of motion of a distant particle,
momentarily stationary, in the coordinate system $x^{\mu }$ of the frame of
reference of the Centre of Mass. Transforming coordinates into this system
the particle would also experience the scalar field acceleration described
in Equation \ref{eq.106} and as the affine connection is defined by 
\begin{equation*}
\Gamma _{\mu \nu }^{\alpha }=\frac{\partial x^{\alpha }}{\partial \xi
^{\beta }}\frac{\partial ^{2}\xi ^{\beta }}{\partial x^{\mu }\partial x^{\nu
}}\text{ ,}
\end{equation*}%
then, if the pressure is negligible, 
\begin{equation}
\frac{d^{2}x^{\alpha }}{d\tau ^{2}}+\Gamma _{\mu \nu }^{\alpha }\frac{%
dx^{\mu }}{d\tau }\frac{dx^{\nu }}{d\tau }=-g^{\alpha \nu }\frac{\kappa
\lambda }{2}\frac{\phi ;_{\nu }}{\phi }\text{ .}  \label{eq.109}
\end{equation}%
Therefore for a slow particle 
\begin{equation}
\frac{d^{2}x^{\alpha }}{d\tau ^{2}}+\Gamma _{00}^{\alpha }\left( \frac{dt}{%
d\tau }\right) ^{2}=-g^{\alpha \nu }\frac{\kappa \lambda }{2}\frac{\phi
;_{\nu }}{\phi }\text{ ,}  \label{eq.110}
\end{equation}%
also 
\begin{equation*}
\frac{d^{2}t}{d\tau ^{2}}=0\text{ ,}\qquad \text{so\qquad }\frac{dt}{d\tau }=%
\sqrt{-g_{00}^{-1}}\text{ which is a constant at }r=r_{1}\text{ .}
\end{equation*}%
So multiplying through by $\left( \frac{d\tau }{dt}\right) ^{2}$ produces 
\begin{equation}
\frac{d^{2}x^{\alpha }}{dt^{2}}+\Gamma _{00}^{\alpha }=+g_{00}\,g^{\alpha
\nu }\frac{\kappa \lambda }{2}\frac{\phi ;_{\nu }}{\phi }\text{ .}
\label{eq.111}
\end{equation}%
Now for a stationary field 
\begin{equation}
\Gamma _{00}^{\alpha }=-\frac{1}{2}g^{\alpha \nu }\frac{\partial g_{00}}{%
dx^{\nu }}  \label{eq.112}
\end{equation}%
and for a weak field, $g_{\alpha \beta }=\eta _{\alpha \beta }+h_{\alpha
\beta }$ , where $\mid h_{\alpha \beta }\mid \ll 1$ and $\eta _{\alpha \beta
}$. The resulting affine connection, linearized in the metric perturbation $%
h_{\alpha \beta }$, becomes in the spherically symmetric case, (as in GR) 
\begin{equation}
\Gamma _{00}^{r}=-\frac{1}{2}\eta ^{rr}\frac{dh_{00}}{dr}  \label{eq.113}
\end{equation}%
so the only non zero component of Equation \ref{eq.111} is 
\begin{equation}
\frac{d^{2}r}{dt^{2}}=\frac{1}{2}\frac{dh_{00}}{dr}+g_{00}\,g^{rr}\frac{%
\kappa \lambda }{2}\frac{1}{\phi }\frac{d\phi }{dr}\text{ .}  \label{eq.114}
\end{equation}%
The general standard form of the metric in both the BD and SCC theories is 
\begin{eqnarray}
d\tau ^{2} &=&\left[ 1-\frac{2G_{m}M}{r}+2\left( 1-\gamma \right) \left( 
\frac{G_{m}M}{r}\right) ^{2}+...\right] dt^{2}  \label{eq.115} \\
&&\ \ \ \ -\left( 1+\frac{2\gamma G_{m}M}{r}+...\right) dr^{2}+r^{2}d\theta
^{2}+r^{2}\sin ^{2}\theta d\varphi ^{2}\text{ ,}  \notag
\end{eqnarray}%
where in both theories $\gamma =\frac{\left( 2-\lambda \right) }{\left(
2+\lambda \right) }$ , therefore 
\begin{equation}
g_{00}=-\left[ 1-\frac{2G_{m}M}{r}+\frac{4\lambda }{\left( 2+\lambda \right) 
}\left( \frac{G_{m}M}{r}\right) ^{2}+...\right] \text{ ,}  \label{eq.116}
\end{equation}%
\begin{equation}
h_{00}=\frac{2G_{m}M}{r}-\frac{4\lambda }{\left( 2+\lambda \right) }\left( 
\frac{G_{m}M}{r}\right) ^{2}+...  \label{eq.117}
\end{equation}%
and 
\begin{equation}
g_{rr}=1+\frac{2\left( 2-\lambda \right) G_{m}M}{\left( 2+\lambda \right) r}%
+...\text{ .}  \label{eq.118}
\end{equation}%
Substituting equations \ref{eq.99}, \ref{eq.116}, \ref{eq.117} and \ref%
{eq.118} in Equation \ref{eq.114} yields 
\begin{eqnarray}
\frac{d^{2}r}{dt^{2}} &=&-\left[ 1-\frac{\kappa \lambda ^{2}}{\left(
2+\lambda \right) }\right] \frac{G_{m}M}{r^{2}}  \label{eq.119} \\
&&\ \ \ \ \ \ \ \ +\left\{ \frac{4\lambda \left( 2+\lambda -2\kappa \lambda
\right) }{\left( 2+\lambda \right) ^{2}}\right\} \frac{\left( G_{m}M\right)
^{2}}{r^{3}}+...  \notag
\end{eqnarray}%
and therefore to first order the total acceleration experienced by a
particle is 
\begin{equation}
\frac{d^{2}r}{dt^{2}}=-\left[ 1-\frac{\kappa \lambda ^{2}}{2+\lambda }\right]
\frac{G_{m}M}{r^{2}}+...\text{ .}  \label{eq.120}
\end{equation}%
However Newtonian gravitational theory defines $G_{N\text{ }}$ by 
\begin{equation}
\frac{d^{2}r}{dt^{2}}=-\frac{G_{N}M}{r^{2}}\text{ ,}  \label{eq.121}
\end{equation}%
therefore the effect of violating the equivalence principle in accordance
with the PMI is that every mass experiences an extra acceleration similar to
Newtonian gravitation and which therefore is confused with it. According to
SCC the Newtonian gravitational constant $G_{N}$, as measured in a Cavendish
type experiment, is a compilation of the effect of the curvature of space
time, with its corresponding $G_{m}$, and the action of the scalar field.
The total Newtonian 'constant' experienced by a mass particle is therefore 
\begin{equation}
G_{N}=\left[ 1-\frac{\kappa \lambda ^{2}}{\left( 2+\lambda \right) }\right]
G_{m}\text{ .}  \label{eq.122}
\end{equation}%
Note that $G_{N}$ and $G_{m}$ refer to the total \textquotedblright
gravitational\textquotedblright\ accelerations experienced in physical
experiments by atomic particles and photons respectively.

\subsubsection{The relationship between $\protect\phi $ and $m$.}

Consider the general Gauss Divergence theorem applied to the gradient of the
Newtonian potential $\Phi _{N}$%
\begin{equation*}
\underset{V}{\iiint }\nabla \Theta \circ dV=\underset{S}{\iint }\Theta \circ
dS\text{ ,}\qquad
\end{equation*}%
\begin{equation*}
\text{put }\underline{\Theta }=\underline{\nabla }\Phi _{N}\quad \text{and
define }\Phi _{N}\text{ by\quad }\nabla ^{2}\Phi _{N}=4\pi G_{N}\rho \quad 
\text{with\quad }\underset{r\rightarrow \infty }{Lim}\Phi _{N}\left(
r\right) =0\text{ ,}
\end{equation*}%
\begin{equation}
\underset{V}{\iiint }\nabla ^{2}\Phi _{N}\,dV=\underset{S}{\iint }\underline{%
\nabla }\Phi _{N}.\underline{dS}\text{ .}  \label{eq.123}
\end{equation}%
In the spherically symmetric One Body case the volume integral on the left
hand side is simply $4\pi GM$ where $M$ is the remote determination of the
total mass of the central body radius $R$. Consider several concentric
external spheres of radius $r_{1}$, $r_{2}$ etc. $\geq R$ centered on the
mass $M$ . As the contributions from the vacuum are zero the volume
integrals over each sphere are equal. 
\begin{equation}
\underset{V_{1}}{\iiint }\nabla ^{2}\Phi _{N}.dV=\underset{V_{2}}{\iiint }%
\nabla ^{2}\Phi _{N}.dV=4\pi GM\text{ .}  \label{eq.124}
\end{equation}%
Therefore observers on the surface of each sphere will have different
determinations of the central mass, which will vary $M\propto m_{i}^{-1}$in
the JF, that is when comparing $M$ to their locally determined atomic masses 
$m_{i}$ by observing the red shift of photons that are emitted from the
central mass with invariant energy. As $GM$ is constant for all $r\geq R$
they will conclude 
\begin{equation*}
G\left( r\right) \propto M^{-1}\left( r\right) \propto m_{i}\left( r\right) 
\text{ .}
\end{equation*}%
But $G(r)=\frac{\psi }{\phi }$ therefore consistency demands 
\begin{equation}
\phi \left( r\right) \propto m_{i}\left( r\right) ^{-1}\text{ .}
\label{eq.125}
\end{equation}%
Integrating the surface integral on the right hand side of equation \ref%
{eq.123} over the sphere at constant $r$ gives, of course, $4\pi r^{2}\nabla
\Phi _{N}\left( r\right) $. (In the standard general form of the metric the
surface area of a sphere is $4\pi r^{2}$. This fact is used both here and
below). The Newtonian potential is defined by the measurement of
acceleration of a 'freeling falling' test particle in a local experiment:%
\begin{equation*}
\frac{d^{2}r}{dt^{2}}=-\nabla \Phi _{N}\left( r\right)
\end{equation*}

Now consider a fixed observer at the CoM in a proper laboratory, where $M$
is constant, who would conclude from equations \ref{eq.123} and \ref{eq.124}
and remembering that $G(r)=\frac{\psi }{\phi }$that 
\begin{equation}
\frac{d^{2}r}{dt^{2}}=-\frac{\psi }{\phi }\frac{M}{r^{2}}\text{ .}
\label{eq.126}
\end{equation}

\subsection{Evaluating $\protect\lambda $, $\protect\kappa $ and $\protect%
\psi $.}

The parameters $\lambda $, $\kappa $, and $\psi $, will be calculated. There
are two methods of calculating the combined gravitational and scalar field
acceleration, one derived from the equations of motion, equation \ref{eq.119}
and the other derived from the definition of the Newtonian potential applied
to Gauss Divergence theorem: equation \ref{eq.126}. Consistency between
these two methods places constraints on the three parameters. Substituting
for $\bar{\omega}=\frac{1}{\lambda }-\frac{3}{2}$ into equation \ref{eq.80}
we obtain the relationship between $G_{m}$ and $G_{0}$: 
\begin{equation}
G_{m}=\left( \frac{2+\lambda }{2}\right) G_{0}\text{ ,}  \label{eq.127}
\end{equation}%
and using this to substitute for $G_{0}$ in equation \ref{eq.119} the
combined gravitational and scalar field acceleration of a free falling
massive body is given by 
\begin{equation}
\frac{d^{2}r}{dt^{2}}=-\left\{ \frac{1}{2}\left( 2+\lambda -\kappa \lambda
^{2}\right) -\left( 2+\lambda -2\kappa \lambda \right) \frac{\lambda G_{0}M}{%
r}+...\right\} \frac{G_{0}M}{r^{2}}\text{ .}  \label{eq.128}
\end{equation}%
But we also have an expression for this combined acceleration from equation %
\ref{eq.126} together with the solution for $\phi $ in equation \ref{eq.98},
expanded for small $\frac{G_{m}M}{r}$. Using equation \ref{eq.127} to
replace $G_{m}$\ with $G_{0}$\ this becomes 
\begin{equation}
\frac{d^{2}r}{dt^{2}}=-\psi \left[ 1-\frac{\lambda G_{0}M}{r}+...\right] 
\frac{G_{0}M}{r^{2}}\text{ .}  \label{eq.129}
\end{equation}%
Comparing coefficients between equations \ref{eq.128} and \ref{eq.129} sets
two conditions on $\lambda $ , $\kappa $, and $\psi $. Consistency between
the coefficients of $\frac{G_{0}M}{r^{2}}$ requires 
\begin{equation}
2+\lambda -\kappa \lambda ^{2}=2\psi  \label{eq.130}
\end{equation}%
and consistency between the coefficients of $\frac{\left( G_{0}M\right) ^{2}%
}{r^{3}}$ requires 
\begin{equation}
2+\lambda -2\kappa \lambda =\psi \text{ .}  \label{eq.131}
\end{equation}

Furthermore we have two solutions for $\phi $; one from the solution to the
scalar field equation in the One Body Problem, and the other from the local
conservation of energy. The first solution, derived from equation \ref{eq.98}
is 
\begin{equation}
\phi =\phi _{0}\exp \left[ \frac{2\lambda }{2+\lambda -\kappa \lambda ^{2}}%
\frac{G_{N}M}{r}\right]  \label{eq.132}
\end{equation}%
the second solution from equations \ref{eq.10} and \ref{eq.125} is 
\begin{equation}
\phi =\phi _{0}\exp \left[ \frac{G_{N}M}{r}\right] \text{ ,}  \label{eq.133}
\end{equation}%
so consistency between these last two solutions, equations \ref{eq.132} and %
\ref{eq.133} sets a third condition on the three parameters 
\begin{equation}
2-\lambda -\kappa \lambda ^{2}=0\text{ .}  \label{eq.134}
\end{equation}%
There are three simultaneous equations \ref{eq.130}, \ref{eq.131} and \ref%
{eq.134} for $\psi $, $\lambda $ and $\kappa $. Their unique solution is 
\begin{equation}
\psi =1\text{ ,}\qquad \lambda =1\qquad \text{and\qquad }\kappa =1\text{ .}
\label{eq.135}
\end{equation}%
Furthermore Equations \ref{eq.125} and \ref{eq.127} give the result 
\begin{equation}
G_{N}=\frac{1}{2}\left( 2+\lambda -\kappa \lambda ^{2}\right) G_{0}=G_{0}=%
\underset{r\rightarrow \infty }{\quad Lim}\frac{1}{\phi \left( r\right) }%
\text{ .}  \label{eq.136}
\end{equation}

Thus $G_{N}$ is the proper value of $\phi ^{-1}$ as measured by atomic
apparatus at infinity, and will be that value determined by physical
apparatus in \textquotedblright Cavendish\textquotedblright\ type
experiments elsewhere.

\subsubsection{The evaluation of $\protect\omega $ and basic relationships}

In conclusion, we have confirmed the findings published earlier (Barber,
2002). Following through carefully the consequences of introducing the
principles of mutual interaction and the local conservation of energy we
have determined that the three parameters introduced into the equations: $%
\lambda $, $\kappa $ and $\psi $ are all unity. The original BD coupling
constant becomes 
\begin{equation}
\varpi =\frac{1}{\lambda }-\frac{3}{2}=-\frac{1}{2}\text{ , \quad }
\label{eq.137}
\end{equation}%
and finally the values of $\lambda $ and $\kappa $ yields from equation \ref%
{eq.42} the SCC coupling constant: 
\begin{equation}
\omega =\frac{1}{\lambda }-\frac{3}{2}-\kappa =-\frac{3}{2}\text{ .}
\label{eq.138}
\end{equation}%
Hence the value $\omega =-\frac{3}{2}$ required in SCC in order to make the
EF of the theory canonical GR is just that value required for consistency
between the two underlying principles of the theory.

With these values the field equations become:

the scalar field equation,%
\begin{equation}
\Box \phi =4\pi T_{M}^{\;},  \label{eq.139a}
\end{equation}

the gravitational field equation,

\begin{eqnarray}
R_{\mu \nu }-\frac{1}{2}g_{\mu \nu }R &=&\frac{8\pi }{\phi }T_{M\mu \nu }-%
\frac{3}{2\phi ^{2}}\left( \nabla _{\mu }\phi \nabla _{\nu }\phi -\frac{1}{2}%
g_{\mu \nu }g^{\alpha \beta }\nabla _{\alpha }\phi \nabla _{\beta }\phi
\right)  \label{eq.140} \\
&&\ \ +\frac{1}{\phi }\left( \nabla _{\mu }\nabla _{\nu }\phi -g_{\mu \nu
}\Box \phi \right) \text{ ,}  \notag
\end{eqnarray}

and the creation equation,

\begin{equation}
\nabla _{\mu }T_{M\nu }^{\;\mu }=\frac{1}{8\pi }\frac{\nabla _{\nu }\phi }{%
\phi }\Box \phi =\frac{1}{2}\frac{\nabla _{\nu }\phi }{\phi }T_{M}\text{.}
\label{eq.141}
\end{equation}

From equation \ref{eq.84} the effective Robertson parameters are 
\begin{equation}
\alpha _{eff}=1\qquad \beta _{eff}=1\qquad \gamma _{eff}=\frac{1}{3}\text{ ,}
\label{eq.142}
\end{equation}%
therefore the standard form of the Schwarzschild metric becomes 
\begin{equation}
d\tau ^{2}=\left( 1-\frac{3G_{N}M}{r}+..\right) dt^{2}-\left( 1+\frac{G_{N}M%
}{r}+..\right) dr^{2}\ \ \ -r^{2}d\theta ^{2}-r^{2}\sin ^{2}\theta d\varphi
^{2}\text{ .}  \label{eq.143}
\end{equation}

The formula for $m_{p}\left( x_{\mu }\right) $ is given by equation \ref%
{eq.10}

\begin{equation}
m_{p}(x^{\mu })=m_{0}\exp [\Phi _{N}\left( x^{\mu }\right) ]\text{ ,}
\label{eq.144}
\end{equation}

in addition from equation \ref{eq.82} we obtain 
\begin{equation}
\phi \left( x_{\mu }\right) =G_{N}^{-1}\exp [-\Phi _{N}\left( x^{\mu
}\right) ]\text{. }  \label{eq.145}
\end{equation}

From equation \ref{eq.127} we have%
\begin{equation}
G_{N}=\frac{2}{3}G_{m}\text{ ,}  \label{eq.145a}
\end{equation}%
so the acceleration of a massive body caused by the curvature of space-time
is $\frac{3}{2}$ the Newtonian gravitational acceleration actually
experienced. However this is compensated by an acceleration caused by the
scalar field of $\frac{1}{2}$ Newtonian gravity in the opposite direction.
This may form the basis for testing the theory in an experiment that asks
the question; "Do photons fall at $\frac{3}{2}$ the rate of freely falling
test particles?"

Finally we have verified the earlier result (Barber, 2002) that in both
equation \ref{eq.128} and \ref{eq.129} if we substitute the values $\lambda
=\kappa =\psi =1$ then we obtain: 
\begin{equation}
\frac{d^{2}r}{dt^{2}}=-\left\{ 1-\frac{G_{N}M}{r}+...\right\} \frac{G_{N}M}{%
r^{2}}\text{ .}  \label{eq.145b}
\end{equation}%
The effect of this non-Newtonian perturbation, adapting the Newtonian
potential to allow for changes in potential energy, was examined in the 2002
paper and was shown to exactly compensate for the effect of the scalar field
on the metric, as indeed the SCC JF conformal equivalence with canonical GR
in its EF would lead us to expect.

\subsection{Experiment and observation}

\subsubsection{The Definitive experiment}

Predictions for the standard experiments that test the trajectories of test
particles and radiation \textit{in vacuo }are all identical to GR, this is
because of the exact compensating nature of the scalar field with its
non-minimal connection to matter. [Note: because of the second order term, $%
\left( \frac{GM}{r}\right) ^{2}$, in equation \ref{eq.119} the equivalence
principle is expermentally violated in E\"{o}tvos type experiments to about
one part in $10^{-17}$\ or three orders of magnitude smaller than present
day experimental sensitivity.(Barber, 2005)] Physical experiments on the
curvature of space-time around a gravitating body, such as the Gravity Probe
B experiment have to be evaluated using the 'true' stress energy tensor and
its field equations, (SCC2) which again are identical to GR \textit{in vacuo}%
. ($\alpha _{true}=\beta _{true}=\gamma _{true}=1$) However in this
formulation of the field equations photons do not travel on geodesics of the
metric and therefore the effective JF has to be used ($\alpha _{eff}=1$, $%
\beta _{eff}=1,$ $\gamma _{eff}=\frac{1}{3}$ with $G_{m}=\frac{3}{2}G_{N}$)
when interpreting observations of light crossing space-time, such as the
gravitational deflection of light.

An experiment will now be briefly described to show the difference between
SCC and GR. Compare the behaviour of light with that of matter in free fall,
the question is whether they fall at the same rate or not. Although the
predictions of the deflection of light by massive bodies are equal in both
theories this does not answer this question as the effects of space
curvature and time dilation are convoluted together in this situation and in
SCC exactly compensate for each other. In order to extract the 'free fall'
acceleration on its own the effects of space curvature and time dilation
have to be separated. To achieve this we note that in SCC, according to
equation \ref{eq.145a}, a\ photon in free fall should descend at $\frac{3}{2}
$ the acceleration of matter. Therefore in free fall a beam of light
travelling a distance $l$ across a solid apparatus is deflected downwards
relative to the apparatus by an amount 
\begin{equation}
\delta =\frac{1}{4}g\left( \frac{l}{c}\right) ^{2}\text{ .}  \label{eq.146}
\end{equation}

As an experiment I have suggested launching into earth orbit an annulus, two
meters in diameter, supporting N (where N \symbol{126}1,000) carefully
aligned small mirrors. A laser beam is then split, one half reflected N
times to be returned and recombined with the other half beam, reflected just
once, to form an interferometer at source. If the experiment is in earth
orbit and the annulus orientated on a fixed star, initially orthogonal to
the orbital plane then the gravitational or acceleration stresses on the
frame, would vanish whereas they would predominate on earth. In such a
space, or free fall, experiment SCC predicts a 2 Angstrom interference
pattern shift with orbital periodicity whereas GR predicts a null result.

\subsubsection{Intriguing observations}

\paragraph{A possible explanation of the Pioneer Anomaly}

The Pioneer Anomaly (PA) has been well documented, (Anderson \textit{et al, }%
1998, 2002) and may be a non-Newtonian real effect that is not explainable
by conventional physics. It is measured as a residual blue Doppler shift on
signals returned back to Earth from the two Pioneer spacecraft and the
effect has been constant and equal for both spacecraft from 10AU - 90AU,
outside Saturn's orbit. The value of the frequency change or clock drift is
equal to: $a_{D}=(2.92\pm 0.44)\times 10^{-18}s^{-1}.$This can be
interpreted as an acceleration (either towards the Sun or the Earth) equal
to $a_{P}=(8.74\pm 1.33)\times 10^{-10}m.s^{-2}$.

It does not show up in the orbital dynamics of the outer planets, which
suggests that it cannot be modelled by a modification in the gravitational
field of the Sun. (Iorio, 2007) Furthermore 'normal physics' from on-board
systematics can so far only explain a maximum of the following possible
sources:

i \qquad\ \ Radio Beam Reaction Force $a_{RB}=(1.10\pm 0.10)\times
10^{-10}m.s^{-2}$.

ii \qquad\ Anisotropic Heat Reflection $a_{AH}=(-0.55\pm 0.55)\times
10^{-10}m.s^{-2}$.

iii \qquad Differential Change of the RTG's Radiant Emissivity

$\qquad \qquad \qquad \qquad \qquad \qquad \qquad \qquad a_{RE}=0.85\times
10^{-10}m.s^{-2}$. .

iv \qquad Constant Electrical Heat Radiation as the Source they was not a
viable explanation.

v \qquad\ Helium Expulsion from the RTGs $a_{HE}=(0.15\pm 0.16)\times
10^{-10}m.s^{-2}$.

vi \qquad Propulsive Mass Expulsion $a_{PME}=\pm 0.56\times 10^{-10}m.s^{-2}$%
. (Turyshev et al. 2010),

In total these various possible sources make a maximum total of $%
a_{N}=(2.1\pm 0.8)\times 10^{-10}m.s^{-2}$ that can be caused by normal
physics leaving at least a minimum anomalous acceleration of $a_{X}=(6.6\pm
2.1)\times 10^{-10}m.s^{-2}$ yet to be explained.

This is equivalent to a minimum Doppler shift or clock drift of

$\qquad \qquad \qquad \qquad \qquad \qquad \qquad \qquad a_{D\text{ }%
residual}=(2.20\pm 0.70)\times 10^{-18}s^{-1}$.

It may be pertinent to note the Hubble parameter in similar units is equal
to: $H=(2.4\pm 0.2)\times 10^{-18}s^{-1}$(using $h=0.73$ with $\pm 10\%$
error bars), which is consistent with that unexplained residual.

In the cosmological solution of the SCC JF atomic masses increase secularly
with 
\begin{equation}
m_{p}=m_{0}\exp (H_{0}t)  \label{eq.147}
\end{equation}%
thus speeding up atomic processes and clocks when compared to orbital
periods, which remain constant in the theory. Therefore intriguingly the
theory predicts a clock drift between ephemeris and atomic clocks of
precisely $H=(2.4\pm \ 0.2)\times \ 10^{-18}s^{-1}$ and this may be the
explanation for the anomaly.

\paragraph{A secular increase of the Earth's rotation rate}

A second anomaly as reviewed by Leslie Morrison and Richard Stephenson
[(Morrison and Stephenson, 1998), (Stephenson, 2003)] arises from the
analysis of the length of the day from ancient eclipse records. It is that
in addition to the tidal contribution there is a long-term component acting
to decrease the length of the day which equals 
\begin{equation*}
\vartriangle \text{T/day/cy}=-6\ \text{x }10^{-4}\text{ sec/day/cy.}
\end{equation*}%
This component, which is consistent with recent measurements made by
artificial satellites, is thought to result from the decrease of the Earths
oblateness following the last ice age. Although this explanation certainly
merits careful consideration, and it is difficult to separate the various
components of the Earth's rotation, it is remarkable that this value $%
\vartriangle $T/day/cy is equal to $H_{0}$ if $H_{0}=67$ $km.sec^{\text{-1}%
}/Mpc$. The question is why should this spinning up of the Earth's rotation
have a natural time scale of the order of the age of the universe rather
than the natural relaxation time of the Earths crust or the periodicity of
the ice ages? This anomaly may be cosmological rather than geophysical in
nature and possibly explained by SCC in which dynamical problems are to be
analysied in the JF. In its cosmological solution atomic masses increase
secularly according to equation \ref{eq.147}, consequently their radii will
shrink (as the Bohr radius is inversely proportional to the mass,) and if
angular momentum $mr^{2}\omega $ is conserved then we have: 
\begin{equation}
m\left( t\right) =m_{0}\exp \left( H_{0}t\right) \text{ and }r(t)=r_{0}\exp
\left( -H_{0}t\right)  \label{eq.147a}
\end{equation}%
\begin{equation}
\text{and if }\frac{d}{dt}\left( mr^{2}\omega \right) =0  \notag
\end{equation}%
\begin{equation}
\text{Then }\frac{\dot{\omega}}{\omega _{0}}=-\left( \frac{\dot{m}}{m_{0}}+2%
\frac{\dot{r}}{r_{0}}\right) =+H_{0}\text{ ,}  \label{eq.147b}
\end{equation}%
therefore solid bodies such as the Earth should spin up when measured in
ephemeris time at a rate equal to the Hubble parameter as indeed may have
already been observed.

\paragraph{A comparison of the temporal and spatial Newtonian potentials of
the metric}

We may compare SCC against GR by casting the Schwarzschild metric as:%
\begin{equation}
d\tau ^{2}=\left[ 1-2\Psi (r)\right] dt^{2}-\left[ 1+2\Theta \left( r\right) %
\right] dr^{2}-r^{2}d\theta ^{2}-r^{2}\sin ^{2}\theta d\varphi ^{2}\text{ .}
\label{eq.148}
\end{equation}

where $\Psi (r)$ and $\Theta \left( r\right) $\ are respectively the
temporal and spatial Newtonian potentials of the theory. They may be
compared by defining 
\begin{equation}
\eta =\frac{\Theta \left( r\right) }{\Psi (r)}  \label{eq.149}
\end{equation}

While it is obvious that for GR $\eta =1$, we have from equation \ref{eq.143}
for SCC $\eta =\frac{1}{3}$. The temporal gravitational potential is
therefore three times larger than the spatial one. It might be possible to
detect such a deviation from GR in appropriate surveys, for example in the
analysis of Large Scale Structure growth.

\section{Cosmological solutions to the field equations}

The solution to the field equations in the cosmological case cast in the JF
have been published earlier. (Barber, 2002) In the JF the cosmological model
is static and eternal with exponentially 'increasing atomic particle
masses', 'shrinking' rulers and 'accelerating' clocks. When transformed into
the EF, with constant particle masses, 'fixed' rulers and 'regular' atomic
clocks, the cosmological model is that of a linearly expanding universe.
Such a model has been described as a 'freely coasting' or a 'Milne'
universe, (but in SCC spatially closed). From Kolb's initial paper onwards
there have been attempts to show that coasting cosmology models could be
concordant with observations and Big Bang nucleosynthesis. [(Kolb, 1989),
(Batra, A. \textit{et al}, 2000), (Gehlaut, S. \textit{et al}, 2003),
(Sethi, G. \textit{et al}., 2005 a \&\ b)]

On the other hand alternative cosmological solutions exist, using the 'true'
stress energy tensor. In this case the field equations become that of the
second SCC theory (Barber, 1982) and their ramifications have been developed
by many authors (see abbreviated reference list below). It is shown below
that with the addition of the principle of the local conservation of energy
the effective JF equations yield a total density parameter $\Omega =1$.

With the mysteries of Dark Matter and Dark Energy unresolved,\ and
continuing problems with Inflation and quantum gravity, it may be that
alternative models such as SCC should be examined more carefully.

\subsection{The JF cosmological equations}

The 'true JF' is to be used when dealing with massive particles and the
'effective JF' is to be used to deal with massless particles such as light
or gravitons. Hence, the 'effective JF' is used to work out cosmological
evolution and then the 'true JF' will be used to calculate the density
parameter.

From the earlier paper, (Barber, 2002), we have the following cosmological
equations for the Robertson-Walker metric and a perfect fluid:

the scalar field equation 
\begin{equation}
\ddot{\phi}+\,3\frac{\dot{\phi}\dot{R}}{R}=4\pi \left( \rho -3p\right) \text{
,}  \label{eq.150}
\end{equation}%
two gravitational equations

\begin{equation}
\left( \frac{\dot{R}}{R}\right) ^{2}+\frac{k}{R^{2}}=+\frac{8\pi \rho }{%
3\phi }-\frac{\dot{\phi}\dot{R}}{\phi R}-\frac{1}{4}\left( \frac{\dot{\phi}}{%
\phi }\right) ^{2}\text{ }  \label{eq.151}
\end{equation}%
and%
\begin{equation}
\frac{\ddot{R}}{R}+\left( \frac{\dot{R}}{R}\right) ^{2}+\frac{k}{R^{2}}=-%
\frac{1}{6}\left( \frac{\ddot{\phi}}{\phi }+3\frac{\dot{\phi}\dot{R}}{\phi R}%
\right) +\frac{1}{4}\left( \frac{\dot{\phi}}{\phi }\right) ^{2}\text{ ,}
\label{eq.152}
\end{equation}%
the creation equation (replacing the conservation equation of GR) 
\begin{equation}
\dot{\rho}\,=-3\frac{\dot{R}}{R}\left( \rho +p\right) +\frac{1}{8\pi }\frac{%
\dot{\phi}}{\phi }\left( \ddot{\phi}+\,3\frac{\dot{\phi}\dot{R}}{R}\right) 
\text{ ,}  \label{eq.153}
\end{equation}%
and the equation of state 
\begin{equation}
p=\sigma \rho \text{ .}  \label{eq.154}
\end{equation}

In the coordinate system of the JF (with time based on the frequency of the
CMB) the universe is static with $R=R_{0}$ a constant, therefore these
equations become:%
\begin{equation}
\ddot{\phi}=4\pi \rho \left( 1-3\sigma \right) \text{ ,}  \label{eq.155}
\end{equation}

\begin{equation}
\frac{k}{R_{0}^{2}}=\frac{8\pi \rho }{3\phi }-\frac{1}{4}\left( \frac{\dot{%
\phi}}{\phi }\right) ^{2}\text{ ,}  \label{eq.156}
\end{equation}

\begin{equation}
\frac{k}{R_{0}^{2}}=-\frac{1}{6}\left( \frac{\ddot{\phi}}{\phi }\right) +%
\frac{1}{4}\left( \frac{\dot{\phi}}{\phi }\right) ^{2}\text{ ,}
\label{eq.157}
\end{equation}

and%
\begin{equation}
\dot{\rho}\,\,=\frac{1}{8\pi }\frac{\dot{\phi}\ddot{\phi}}{\phi }.
\label{eq.158aa}
\end{equation}%
Add\ equations \ref{eq.156} and \ref{eq.157} and use equation \ref{eq.155}
to eliminate $\rho $\ we obtain%
\begin{equation}
\frac{k}{R_{0}^{2}}=\frac{\left( 1+\sigma \right) }{4\left( 1-3\sigma
\right) }\left( \frac{\ddot{\phi}}{\phi }\right) \text{,}  \label{eq.158b}
\end{equation}

The set of equations in the $k=0$ case leads to the solution 
\begin{equation}
\phi =\phi _{0}\left( \frac{t_{0}}{t}\right) ^{2}\text{ with }\sigma =-1%
\text{.}  \label{eq.158ba}
\end{equation}

However, if we consider a universe with matter (baryonic and dark), and
false energy ($\sigma _{fe}=-1$) but, in the present epoch, negligible
electro-magnetic radiation and matter pressure, then the resultant equation
of state%
\begin{equation*}
p_{fe}=-\left( \rho _{m}+\rho _{fe}\right) \text{,}
\end{equation*}

leads to $\rho _{m}=0$, i.e. an empty universe. A realistic solution of
equation \ref{eq.158b} has to have a non-zero $k$, therefore, with its left
hand side being a non-zero constant, the solution is 
\begin{equation*}
\phi =\phi _{0}\exp \left[ H\left( t-t_{0}\right) \right] \text{,}
\end{equation*}

where $H$\ is some arbitrary constant - calculated in the 2002 paper to be
the Hubble parameter as measured in the present epoch, $H_{0}$ thus%
\begin{equation}
\phi =\phi _{0}\exp \left( H_{0}t\right)  \label{eq.158c}
\end{equation}

Now eliminate $\ddot{\phi}$\ from equations \ref{eq.155} and \ref{eq.158aa} 
\begin{equation*}
\frac{\dot{\rho}}{\rho }=\frac{1}{2}\frac{\dot{\phi}}{\phi }\left( 1-3\sigma
\right)
\end{equation*}

Integrating w.r.t. $t$ between the limits $t_{0}$\ and $t$ we obtain%
\begin{equation}
\rho =\rho _{0}\left( \frac{\phi }{\phi _{0}}\right) ^{\frac{1}{2}\left(
1-3\sigma \right) }\text{,}  \label{eq.159}
\end{equation}

and eliminating $\frac{k}{R^{2}}$\ between equations \ref{eq.156} and \ref%
{eq.157} and then using equation \ref{eq.155} to eliminate $\rho $

results in%
\begin{equation*}
\left( 5-3\sigma \right) \frac{\ddot{\phi}}{\phi }=3\left( 1-3\sigma \right)
\left( \frac{\dot{\phi}}{\phi }\right) ^{2}
\end{equation*}

substituting the solution from equation \ref{eq.158c} we are left with%
\begin{equation*}
\left( 5-3\sigma \right) H_{0}^{2}=3\left( 1-3\sigma \right) H_{0}^{2}\text{,%
}
\end{equation*}

so, in the effective JF, the cosmological equation of state is determined by
the scalar, gravitational and creation field equations to be 
\begin{equation}
\sigma =-\frac{1}{3}\text{.}  \label{eq.160}
\end{equation}

Equation \ref{eq.158b} becomes%
\begin{equation*}
\frac{k}{R_{0}^{2}}=\frac{1}{12}H_{0}^{2}\text{,}
\end{equation*}

as $H_{0}^{2}$\ and $R_{0}^{2}$\ are positive definite, therefore $k=+1$ so
the universe is closed with a scale length%
\begin{equation}
R_{0}=+\sqrt{12}H_{0}^{-1}\simeq 47G.lys.  \label{eq.161}
\end{equation}

Substituting this equation of state in the 'effective JF' the following
cosmological relationships were calculated in the 2002 paper:

\begin{equation*}
\phi =\phi _{0}\exp \left( H_{0}t\right) \text{, where }\phi _{0}=G_{N}^{-1}%
\text{ and }t=0\text{ is the present,}
\end{equation*}

and with the caveat that the 'effective JF' is not appropriate for massive
particles we also have 
\begin{equation*}
\rho _{eff}=\rho _{0}\exp \left( H_{0}t\right) ,
\end{equation*}
from which, with equation \ref{eq.155}, we derive 
\begin{equation*}
\Omega _{eff}=\frac{1}{3},
\end{equation*}
and finally%
\begin{equation}
m_{eff}=m_{0}\exp \left( H_{0}t\right) \text{, where m}_{0}\text{\ is a
particle mass in the present.}  \label{eq.162}
\end{equation}

\subsection{\protect\bigskip The EF Cosmological Equations}

The dynamical evolution of the universe, determined by gravitational and
scalar fields, has to be calculated in the 'effective JF', however the
density parameter of the universe has to be calculated in the 'true JF' in
which the scalar field becomes 'invisible'. In an early paper, (Soleng,
1987), following (Pimentel, 1985), the gravitational equations for the
Robertson-Walker metric and a perfect fluid and with $\lambda =1$ are given
as:

the density gravitational equation

\begin{equation}
\left( \frac{\dot{R}}{R}\right) ^{2}+\frac{k}{R^{2}}=+\frac{8\pi \rho }{%
3\phi }\text{ }  \label{eq.164}
\end{equation}

and the pressure gravitational equation

\begin{equation}
2\frac{\ddot{R}}{R}+\left( \frac{\dot{R}}{R}\right) ^{2}+\frac{k}{R^{2}}=-%
\frac{8\pi p}{\phi }\text{ ,}  \label{eq.165}
\end{equation}

with the equation of state 
\begin{equation}
p=\sigma \rho \text{ .}  \label{eq.167}
\end{equation}

If we subtract equation \ref{eq.164} from \ref{eq.165} to obtain%
\begin{equation}
2\frac{\ddot{R}}{R}=-\frac{8\pi \left( \rho +3p\right) }{3\phi }\text{ ,}
\label{eq.168}
\end{equation}

and use the above equation of state with $\sigma =-\frac{1}{3}$ then $\ddot{R%
}=0$ and the gravitational field equations are consistent with either a
linearly expanding universe or, a static one. 
\begin{equation}
R(t)=t\text{, or }R(t)=t_{0}\text{. (note }c=1\text{)}  \label{eq.169}
\end{equation}

In this fnal version of the theory the JF, in either 'effective' or 'true'
forms, includes the local conservation of energy as an additional principle
to be added to the 1982 theory. With time therefore measured by a photon
sampled from the CMB as the standard unit of measurement so 'light-rulers'
expand with the universe, the solution by definition must be the static $%
R(t)=t_{0}$.

\bigskip As $k=+1$ equation \ref{eq.164} becomes%
\begin{equation}
\frac{1}{t_{0}^{2}}=+\frac{8\pi \rho }{3\phi }\text{,}  \label{eq.170}
\end{equation}

with $t_{0}^{-1}=H_{0}$ and $\phi =G_{N}^{-1}$ \ref{eq.170} is simply the
equation for critical density,%
\begin{equation*}
\frac{8\pi G_{N}\rho }{3H_{0}^{2}}=1\text{.}
\end{equation*}

In this 'true JF', used to measure massive, but not massless, particles the
density of the universe is the critical density and%
\begin{equation}
\Omega _{true}=1  \label{eq.171}
\end{equation}

\section{Conclusions}

The 2002 version of the theory has been corrected to use the 'true' form of
the stress-energy tensor to evaluate experiments and observations dealing
with matter and the 'effective' form of the stress-energy tensor to
interpret those dealing with light. With this correction the theory
correctly predicts the geodetic precession measurement of the Gravity Probe
B experiment, which the 2002 version did not. However that theory, and the
present version, are concordant with all other tests of GR and two further
experiments may resolve this degeneracy. Furthermore the theory offers an
explanation for a real Pioneer Anomaly and also for hints of some other
non-GR anomalies. On the one hand, in the EF the universe is seen to expand
linearly from a Big Bang thus resolving the smoothness and density problems
without the need for Inflation, furthermore papers and eprints examining
primordial nucleosynthesis in such a coasting cosmology suggest the baryon
density would be much higher and might explain Dark Matter as being baryonic
in nature, however what form this baryonic dark matter takes remains an
unanswered question. On the other hand, in the 'effective' Jordan conformal
frame, in which the unit of time is measured by a photon sampled from the
peak of the CMB, the universe is closed and static with masses increasing
exponentially with time, causing solid rulers to shrink and atomic clocks to
accelerate in the same manner. The moment of the Big Bang itself is
projected into the infinite past, thereby avoiding philosophical problems
concerned with the concept of 'an origin'. By using the true form of the JF
the total density parameter $\Omega $\ is determined to be unity.

\section{References}

\begin{itemize}
\item 
\begin{enumerate}
\item Anderson, J.D., Laing, P., Lau, E., Liu, A., Nieto, M.M., Turyshev,
S.G., Phys.Rev.Lett. 81 2858-2861. Indication, from Pioneer 10/11, Galileo,
and Ulysses Data, of an Apparent Anomalous, Weak, Long-Range Acceleration.
1998

\item Anderson, J.D., Laing, P., Lau, E., Liu, A., Nieto, M.M., Turyshev,
S.G., Phys.Rev.D65:082004. Study of the anomalous acceleration of Pioneer 10
and 11. 2002

\item Barber, G.A., Gen Relativ Gravit. 14, 117. On Two Self Creation
Cosmologies. 1982

\item Barber, G.A., Astrophysics and Space Science 282, 683-730. A New Self
Creation Cosmology. 2002

\item Barber, G.A., New Developments in Quantum Cosmology Research. Horizons
in World Physics, Volume 247, pages 155 -- 184, Nova Science Publishers,
Inc. New York. Self Creation Cosmology - An Alternative Gravitational
Theory. (eprint arXiv:gr-qc/0405094.) 2005

\item Barber, G.A., Astrophysics and Space Science 305, 169-176. Resolving
the Degeneracy: Experimental tests of the new self creation cosmolgy and a
hererodox prediction for Gravity Probe B. 2006

\item Batra, A., Lohiya, D., Mahajan, S., Mukherjee, A., International
Journal of Modern Physics D, Volume: 9, Issue: 6 pp. 757-773.
Nucleosynthesis in a universe with a linearly evolving scale factor. 2000

\item Brans, C.H. \& Dicke, R.H., Phys. Rev. 124, 925.Mach's Principle and a
Relativistic Theory of Gravitation. 1961

\item Brans, C.H., Gen Relativ Gravit.19, 949. Consistency of field
equations in self-creation cosmologies. 1987

\item Brans, C.H., arXiv:gr-qc/9705069. Gravity and the Tenacious Scalar
Field - contribution to Festschrift volume for Englebert Sch\"{u}cking.
:1997.

\item Dicke, R..H. Phys. Rev. 125, 2163. Mach's Principle and Invariance
under Transformation of Units. 1962

\item Gehlaut, S., Kumar, P., Geetanjali, Lohiya, D., eprint
arXiv:astro-ph/0306448. A Concordant "Freely Coasting Cosmology". 2003

\item Iorio, L., Found.Phys.D37:897-918, Can the Pioneer anomaly be of
gravitational origin? A phenomenological answer. (eprint
http://arxiv.org/abs/gr-qc/0610050.) 2007

\item Kolb, E.W., Astrophysical Journal, Part 1 (ISSN 0004-637X), vol. 344,
p. 543-550. A coasting cosmology. 1989

\item Magnano, G. \& Sokolowski, L.M., Phys. Rev D 50,8. Physical
equivalence between nonlinear gravity theories and a general-relativistic
self-gravitating scalar field. 1994

\item Milne, E.A., Relativity, Gravitation and World Structure. Clarendon
Press: Oxford. 1935

\item Milne, E.A., Kinematic Relativity a sequel to Relativity, Gravitation
and World Structure. Clarendon Press: Oxford. 1948

\item Morrison, L. Stephenson, R. Astronomy \& Geophysics, Vol. 39, Issue 5,
p.8. The sands of time. 1998

\item Nordstr\"{o}m, G., Annalen der Physik, 40, 856. 1913

\item Quiros, I., arXiv:gr-qc/9904004. Conformal classes of Brans-Dicke
gravity. 1999

\item Quiros, I., arXiv:gr-qc/9905071. Dual geometries and spacetime
singularities. :2001

\item Santiago, D.I. \& Silbergleit, A.S. arXiv:gr-qc/9904003. On the
Energy-Momentum Tensor of the Scalar Field in Scalar-Tensor Theories of
Gravity. 2000,

\item Sethi, G., Kumar, P., Pandey, S., Lohiya, D., eprint
arXiv:astro-ph/0502370. A case for nucleosynthesis in slowly evolving
models. 2005 (a)

\item Sethi, G., Dev, A., Deepak, J., Physics Letters B, Volume 624, Issue
3-4, p. 135-140. eprint arXiv:astro-ph/0506255. Cosmological constraints on
a power law universe. 2005 (b)

\item Stephenson, F.R., Astronomy \& Geophysics Vol 44 issue 2, pages 2.22 -
2.27. Historical eclipses and Earth's rotation. 2003

\item Turyshev, S.G., Toth, V.T., Living Rev. Relativity 13, 4.The Pioneer
Anomaly. eprint http://arxiv.org/abs/1001.3686. 2010

\item Weinberg, S., Gravitation and Cosmology, Principles and Applications
of the General Theory of Relativity. Wiley. 1972

\item Weyl, H., \textquotedblright Gravitation und
Electriticitat\textquotedblright . Sitzungsberichte der Preussichen Akad. d.
Wissenschaften. (English translation, 1923, in Dover Publications.
\textquotedblright The Principle of Relativity\textquotedblright . 1918
\end{enumerate}
\end{itemize}

\subsubsection{References to papers and eprints developing SCC2 cosmological
solutions}

(A selection listed in reverse chronological order)

\begin{enumerate}
\item Adhav, K. S.; Nimkar, A. S.; Mete, V. G.; Dawande, M. V.,
International Journal of Theoretical Physics, Volume 49, Issue 5,
pp.1127-1132. Axially Symmetric Bianchi Type-I Model with Massless Scalar
Field and Cosmic Strings in Barber's Self-Creation Cosmology. 2010

\item Rathore G.S. \& Mandawat K., Adv. Studies Theor. Phys., Vol. 4, no. 5,
213 -- 224. Five Dimensional Perfect Fluid Cosmological Models in Barber's
Second Self-Creation Theory.: 2010

\item Aghmohammadi, A.; Saaidi, Kh.; M. R. Abolhassani ; Vajdi, A., Physica
Scripta, Volume 80, Issue 6, pp. 065008, Standard cosmological evolution in
the f(R) model to Kaluza-Klein cosmology. 2009

\item Dabrowski\_M. , Garecki J. \& Blaschke D.B. eprint arXiv:0806.2683v3.
Conformal transformations and conformal invariance in gravitation. 2009

\item Jain, V.C., Yadau, M.K. \& Mishra, P.K., International Journal of
Theoretical Physics 48,8. Bianchi Type-I Cosmological Model with varying $%
\Lambda $ Term in Self-Creation Theory of Gravitation. 2009

\item Pradhan, Anirudh; Agarwal, Shilpi; \& Singh, G. P., International
Journal of Theoretical Physics, Volume 48, Issue 1, pp.158-166. LRS Bianchi
Type-I Universe in Barber's Second Self Creation Theory.2009

\item Rao V. U. M., \& Vinutha T., Astrophysics and Space Science Online
First. Plane symmetric string cosmological models in self-creation theory of
gravitation.: 2009

\item Adhav, K.S., Nimkar, A.S. \& Dawande, M.V., International Journal of
Theoretical Physics, Online First. Axially Symmetric Cosmological Micro
Model in Barber's Modified Theory of Einstein General Relativity.2008

\item Katore, S. D. Rane R. S. and Kurkure V. B , Astrophysics and Space
Science 315, pgs 347-352. Plane symmetric cosmological models with negative
constant deceleration parameter in self creation theory. 2008

\item Pradhan, A., Agarwal, S., \& Singh, G. P., International Journal of
Theoretical Physics, LRS Bianchi Type-I Universe in Barber's Second Self
Creation theory. 2008

\item Rao V. U. M., Vijaya Santhi M. \& Vinutha T., Astrophysics and Space
Science 314, 1-3. Exact Bianchi type II, VIII and IX string cosmological
models in Saez-Ballester theory of gravitation. :2008

\item Rao V. U. M., Vijaya Santhi M. \& Vinutha T., Astrophysics and Space
Science 317, 1-2 Exact Bianchi type II, VIII and IX string cosmological
models in General Relativity and self-creation theory of gravitation. :2008

\item Reddy, D. R. K. \& Naidu, R. L., International Journal of Theoretical
Physics, Online First, Kaluza-Klein Cosmological Model in Self-Creation
Cosmology. 2008

\item Singh J. P. , Tiwari R. K. \& Kumar, Sushil, Astrophysics and Space
Science 314, 1-3. Bianchi type-I models in self-creation cosmology with
constant deceleration parameter. 2008

\item Tiwari, R.K. Asteroids, Comets, Meteors (2008). Bianchi Type-V models
in Self-Creation Cosmology with constant deceleration parameter 2008

\item Venkateswarlu, R.; Rao, V. U. M.\& Pavan Kumar, K., International
Journal of Theoretical Physics, 47, 3. String Cosmological Solutions in
Self-Creation Theory of Gravitation. 2008

\item Blaschke, D. \& Dabrowski, M: eprint arXiv:hep-th/0407078. Conformal
relativity versus Brans-Dicke and superstring theories. 2004

\item Mohanty, G. \& Mishra, B., Astrophysics and Space Science Volume 281,
Number 3, 577-583, . Vacuum cosmological models in Einstein and Barber
theories. 2002

\item Panigrahi, U.K \& Sahu, R.C., National Academy Science Letters (Natl.
Acad. Sci. Lett.) ISSN 0250-541X. Plane symmetric mesonic stiff fluid models
in self creation cosmology. 2002

\item Abdussattar \& Vishwakarma, R. G., Classical and Quantum Gravity 14 pp
945-953. Some FRW models with variable G and $\Lambda $. 1997

\item Abdel-Rahman, A.M.M.,Astrophysics and Space Science 189, 1.
Singularity-free self-creation cosmology. 1992

\item Tiwari, S.C. Phy. Lett. A, 142, 8-9, p. 460-464. Scalar field in
gravitational theory. 1990

\item Wolf, C., Astronomische Nachrichten 309, 3 pgs.173-175. Higher order
curvature terms in theories with creation. 1988

\item Soleng, Harald H., Astrophysics and Space Science 138, 1. A note on
vacuum self-creation cosmological models. 1987a

\item Soleng, Harald H., Astrophysics and Space Science 139, 1.
Self-creation cosmological solutions. 1987b

\item Pimentel, L.O. Astrophysics and Space Science 116, 2. Exact
self-creation cosmological solutions. 1985
\end{enumerate}

\end{document}